\documentclass[twocolumn,twocolappendix]{aastex63}
\usepackage{natbib,aas_macros,amsmath}
\usepackage{natbib}
\usepackage{longtable}

\usepackage{multirow,color}

\def\tcb{\textcolor{blue}}

\newcommand{\oiii}{[O\,{\sc iii}]}

\newcommand{\cii}{[C\,{\sc ii}]}
\newcommand{\nii}{[N\,{\sc ii}]}

\newcommand{\targbg}{RXCJ0600-$z$6}

\received{2020 October 14}
\revised{2020 December 14} 
\accepted{2020 December 29}
\submitjournal{ApJ in press}
\shorttitle{Strongly lensed \cii\ line from a Sub-$L^{*}$ Lyman-break galaxy at $z=6.07$19}
\shortauthors{Fujimoto et al.}

\begin{document}

\title{
ALMA Lensing Cluster Survey: \\
Bright \cii\ 158 $\mu$m Lines from a Multiply Imaged Sub-$L^{\star}$ Galaxy at $z=6.0719$
}

\correspondingauthor{Seiji Fujimoto}
\email{fujimoto@nbi.ku.dk}

\author[0000-0001-7201-5066]{Seiji Fujimoto}
\affiliation{
Cosmic Dawn Center (DAWN), Jagtvej 128, DK2200 Copenhagen N, Denmark
}
\affiliation{
Niels Bohr Institute, University of Copenhagen, Lyngbyvej 2, DK2100 Copenhagen \O, Denmark
}

\author{Masamune Oguri}
\affiliation{
Research Center for the Early Universe, Graduate School of Science, The University of Tokyo, 7-3-1 Hongo, Bunkyo-ku, Tokyo 113-0033, Japan
}
\affiliation{
Department of Physics, The University of Tokyo, 7-3-1 Hongo, Bunkyo-ku, Tokyo 113-0033, Japan
}
\affiliation{
Kavli Institute for the Physics and Mathematics of the Universe (WPI), The University of Tokyo, 5-1-5 Kashiwanoha, Kashiwa-shi,
Chiba, 277-8583, Japan
}

\author{Gabriel Brammer}
\affiliation{
Cosmic Dawn Center (DAWN), Jagtvej 128, DK2200 Copenhagen N, Denmark
}
\affiliation{
Niels Bohr Institute, University of Copenhagen, Lyngbyvej 2, DK2100 Copenhagen \O, Denmark
}

\author{Yuki Yoshimura}
\affiliation{
Department of Astronomy, Graduate School of Science, The University of Tokyo, 7-3-1 Hongo, Bunkyo-ku, Tokyo 113-0033, Japan
}
\affiliation{
Institute of Astronomy, Graduate School of Science, The University of Tokyo, 2-21-1 Osawa, Mitaka, Tokyo 181-0015, Japan
}

\author{Nicolas Laporte}
\affiliation{
Kavli Institute for Cosmology, University of Cambridge, Madingley Road, Cambridge CB3 0HA, UK
}
\affiliation{
Cavendish Laboratory, University of Cambridge, 19 JJ Thomson Avenue, Cambridge CB3 0HE, UK
}

\author{Jorge Gonz\'alez-L\'opez}
\affiliation{
N\'ucleo de Astronom\'ia de la Facultad de Ingenier\'ia y Ciencias, Universidad Diego Portales, Av. Ejército Libertador 441, Santiago, Chile
}
\affiliation{
Las Campanas Observatory, Carnegie Institution of Washington, Casilla 601, La Serena, Chile
}

\author{Gabriel B. Caminha}
\affiliation{
Kapteyn Astronomical Institute, University of Groningen, Postbus 800, 9700 AV Groningen, The Netherlands
}

\author{Kotaro Kohno}
\affiliation{
Institute of Astronomy, Graduate School of Science, The University of Tokyo, 2-21-1 Osawa, Mitaka, Tokyo 181-0015, Japan
}
\affiliation{
Research Center for the Early Universe, School of Science, The University of Tokyo, 7-3-1 Hongo, Bunkyo-ku, Tokyo 113-0033, Japan
}

\author{Adi Zitrin}
\affiliation{
Physics Department, Ben-Gurion University of the Negev, P.O. Box 653, Be’er-sheva 8410501, Israel
}

\author{Johan Richard}
\affiliation{
Univ Lyon, Univ Lyon1, Ens de Lyon, CNRS, Centre de Recherche Astrophysique de Lyon UMR5574, F-69230, Saint-Genis-Laval,France
}

\author{Masami Ouchi}
\affiliation{
National Astronomical Observatory of Japan, 2-21-1 Osawa, Mitaka, Tokyo 181-8588, Japan
}
\affiliation{
Institute for Cosmic Ray Research, The University of Tokyo, 5-1-5 Kashiwanoha, Kashiwa, Chiba 277–8582, Japan
}
\affiliation{
Kavli Institute for the Physics and Mathematics of the Universe (WPI), The University of Tokyo, 5-1-5 Kashiwanoha, Kashiwa-shi, Chiba, 277-8583, Japan
}

\author{Franz E. Bauer}
\affiliation{
Instituto de Astrofısica, Facultad de Fısica, Pontificia Universidad Catolica de Chile Av. Vicuna Mackenna 4860, 782-0436 Macul,Santiago, Chile
}
\affiliation{
Millennium Institute of Astrophysics (MAS), Nuncio Monse nor Santero Sanz 100, Providencia, Santiago, Chile
}

\author{Ian Smail}
\affiliation{
Centre for Extragalactic Astronomy, Department of Physics, Durham University, South Road, Durham DH1 3LE, UK
}

\author{Bunyo Hatsukade}
\affiliation{
Institute of Astronomy, Graduate School of Science, The University of Tokyo, 2-21-1 Osawa, Mitaka, Tokyo 181-0015, Japan
}

\author{Yoshiaki Ono}
\affiliation{
Institute for Cosmic Ray Research, The University of Tokyo, 5-1-5 Kashiwanoha, Kashiwa, Chiba 277–8582, Japan
}

\author{Vasily Kokorev}
\affiliation{
Cosmic Dawn Center (DAWN), Jagtvej 128, DK2200 Copenhagen N, Denmark
}
\affiliation{
Niels Bohr Institute, University of Copenhagen, Lyngbyvej 2, DK2100 Copenhagen \O, Denmark
}

\author{Hideki Umehata}
\affiliation{
RIKEN Cluster for Pioneering Research, 2-1 Hirosawa, Wako, Saitama 351-0198, Japan
}
\affiliation{
Institute of Astronomy, Graduate School of Science, The University of Tokyo, 2-21-1 Osawa, Mitaka, Tokyo 181-0015, Japan
}

\author{Daniel Schaerer}
\affiliation{
Observatoire de Gen\`eve, Universit\'e de Gen\`eve, 51 Ch. des Maillettes, 1290 Versoix, Switzerland
}
\affiliation{
CNRS, IRAP, 14 Avenue E. Belin, 31400 Toulouse, France
}

\author{Kirsten Knudsen}
\affiliation{
Department of Space, Earth and Environment, Chalmers University of Technology, Onsala Space Observatory, SE-43992 Onsala, Sweden
}

\author{Fengwu Sun}
\affiliation{
Steward Observatory, University of Arizona, 933 N. Cherry Ave, Tucson, AZ 85721, USA
}

\author{Georgios Magdis}
\affiliation{
Cosmic Dawn Center (DAWN), Jagtvej 128, DK2200 Copenhagen N, Denmark
}
\affiliation{
Niels Bohr Institute, University of Copenhagen, Lyngbyvej 2, DK2100 Copenhagen \O, Denmark
}
\affiliation{
DTU-Space, Technical University of Denmark, Elektrovej 327, DK2800 Kgs. Lyngby, Denmark
}

\author{Francesco Valentino}
\affiliation{
Cosmic Dawn Center (DAWN), Jagtvej 128, DK2200 Copenhagen N, Denmark
}
\affiliation{
Niels Bohr Institute, University of Copenhagen, Lyngbyvej 2, DK2100 Copenhagen \O, Denmark
}

\author{Yiping Ao}
\affiliation{
Purple Mountain Observatory and Key Laboratory for Radio Astronomy, Chinese Academy of Sciences, Nanjing, China
}

\author{Sune Toft}
\affiliation{
Cosmic Dawn Center (DAWN), Jagtvej 128, DK2200 Copenhagen N, Denmark
}
\affiliation{
Niels Bohr Institute, University of Copenhagen, Lyngbyvej 2, DK2100 Copenhagen \O, Denmark
}

\author{Miroslava Dessauges-Zavadsky}
\affiliation{
Observatoire de Gen\`eve, Universit\'e de Gen\`eve, 51 Ch. des Maillettes, 1290 Versoix, Switzerland
}

\author{Kazuhiro Shimasaku}
\affiliation{
Department of Astronomy, Graduate School of Science, The University of Tokyo, 7-3-1 Hongo, Bunkyo-ku, Tokyo 113-0033, Japan
}
\affiliation{
Research Center for the Early Universe, Graduate School of Science, The University of Tokyo, 7-3-1 Hongo, Bunkyo-ku, Tokyo 113-0033, Japan
}

\author{Karina Caputi}
\affiliation{
Kapteyn Astronomical Institute, University of Groningen, P.O. Box 800, 9700AV Groningen, The Netherlands
}
\affiliation{
Cosmic Dawn Center (DAWN), Jagtvej 128, DK2200 Copenhagen N, Denmark
}

\author{Haruka Kusakabe}
\affiliation{
Observatoire de Gen\`eve, Universit\'e de Gen\`eve, 51 Ch. des Maillettes, 1290 Versoix, Switzerland
}

\author{Kana Morokuma-Matsui}
\affiliation{
Institute of Astronomy, Graduate School of Science, The University of Tokyo, 2-21-1 Osawa, Mitaka, Tokyo 181-0015, Japan
}

\author{Kikuchihara Shotaro}
\affiliation{
Department of Astronomy, Graduate School of Science, The University of Tokyo, 7-3-1 Hongo, Bunkyo-ku, Tokyo 113-0033, Japan
}
\affiliation{
Institute for Cosmic Ray Research, The University of Tokyo, 5-1-5 Kashiwanoha, Kashiwa, Chiba 277–8582, Japan
}

\author{Eiichi Egami}
\affiliation{
Steward Observatory, University of Arizona, 933 N. Cherry Ave, Tucson, AZ 85721, USA
}

\author{Minju M. Lee}
\affiliation{
Max-Planck-Institut f\:ur Extraterrestrische Physik (MPE), Giessenbachstr., D-85748, Garching, Germany.
}

\author{Timothy Rawle}
\affiliation{
European Space Agency (ESA), ESA Office, Space Telescope Science Institute, 3700 San Martin Drive, Baltimore, MD 21218, USA
}

\author{Daniel Espada}
\affiliation{
SKA Organisation, Lower Withington, Macclesfield, Cheshire SK11 9DL, UK
}

\def\apj{ApJ}%
\def\apjl{ApJ}%
\def\apjs{ApJS}%

\def\rme{\rm e}
\def\rmstar{\rm star}
\def\rmFIR{\rm FIR}
\def\itHubble{\it Hubble}
\def\rmyr{\rm yr}

\begin{abstract}
We present bright \cii\ 158 $\mu$m line detections from a strongly magnified and multiply-imaged  ($\mu\sim20$--160) sub--$L^{*}$ ($M_{\rm UV}$ = $-19.75^{+0.55}_{-0.44}$) Lyman-break galaxy (LBG) at $z=6.0719\pm0.0004$ from the ALMA Lensing Cluster Survey (ALCS). 
Emission lines are identified at 268.7 GHz at $\geq$ 8$\sigma$ exactly at positions of two multiple images of the LBG behind the massive galaxy cluster RXCJ0600$-$2007. 
Our lens models, updated with the latest spectroscopy from VLT/MUSE, indicate that a sub region of the LBG crosses the caustic and is lensed into a long ($\sim6''$) arc with a local magnification of $\mu\sim 160$, for which the \cii\ line is also significantly detected.
The source-plane reconstruction resolves the interstellar medium (ISM) structure, showing that the \cii\ line is co-spatial with the rest-frame UV continuum at the scale of $\sim$300 pc. 
The \cii\ line properties suggest that the LBG is a rotation-dominated system whose velocity gradient explains a slight difference of redshifts between the whole LBG and its sub region. 
The star formation rate (SFR)--$L_{\rm [CII]}$ relations from the sub to the whole regions of the LBG are consistent with those of local galaxies.
We evaluate the lower limit of the faint-end of the \cii\ luminosity function at $z=6$, and find that it is consistent with predictions from semi analytical models and from the local SFR--$L_{\rm [CII]}$ relation with a SFR function at $z=6$. 
These results imply that the local SFR--$L_{\rm [CII]}$ relation is universal for a wide range of scales including the spatially resolved ISM, the whole region of galaxy, and the cosmic scale, even in the epoch of reionization.
\end{abstract}

\section{Introduction}\label{sec:intro}

Galaxy evolution is regulated by several key mechanisms in the interstellar medium (ISM) such as disk formation, stellar and active galactic nuclei (AGN) feedback, mass building via star formation and galaxy mergers, and clump formations through disk instabilities. 
Resolving the ISM structure to study local physical properties in high-redshift galaxies is thus essential in order to understand the initial phase of galaxy formation and evolution. 

During the past decades, hundreds of star-forming galaxies at $z>6$ have been spectroscopically identified mainly with Ly$\alpha$ lines \citep[e.g.,][]{iye2006,vanzella2011,pentericci2011,pentericci2014,pentericci2018,shibuya2012,shibuya2018,ono2012,ono2018,finkelstein2013,oesch2015,oesch2016,stark2017,higuchi2019}. 
The Atacama Large Millimeter/submillimeter Array (ALMA) offers a rest-frame far-infrared (FIR) spectroscopic window for these $z>6$ galaxies, especially with bright fine-structure lines of \cii\ 158 $\mu$m and \oiii\ 88 $\mu$m \citep[e.g.,][]{maiolino2015,inoue2016,pentericci2016,knudsen2016,matthee2017,matthee2019,carniani2018b,smit2018,bowler2018,hashimoto2018,hashimoto2019,tamura2019,fujimoto2019,bakx2020}. 
Since heavy elements produced in stars are returned into the ISM, 
the metal gas properties traced by the fine-structure lines are good probes of the star-formation history and related physical mechanisms \citep{maiolino2019}. 
In fact, recent ALMA spatial and kinematic \cii-line studies identify signatures of some key mechanisms, including disk rotations \citep[e.g.,][]{jones2017b,smit2018}, galaxy mergers \citep[e.g.,][]{hashimoto2019,lefevre2020}, and outflows \citep[e.g.,][]{gallerani2018,spilker2018,fujimoto2019,fujimoto2020b,ginolfi2020}. 
In conjunction with other fine-structure lines of \oiii\ and \nii, recent ALMA observations also allow us to perform multiple line diagnostics to  constrain the dominant ionization state of the ISM gas \citep[e.g.,][]{inoue2016,pavesi2016,laporte2019, novak2019, harikane2020}. 

There are several challenges related to the FIR spectroscopy. 
The first is sensitivity. 
While ALMA is the most sensitive mm/submm telescope and yielding a large number of new findings about high-redshift galaxies, 
the detection of FIR fine-structure lines from abundant, typical galaxies 
remains challenging.
For example, to observe a \cii\ line of  $\sim1\times$ 10$^{8}\,L_{\odot}$ from $z=6$, about 2-hour observing time is required\footnote{Based on CASA Observing Tool calculations to detect the \cii\ line of 1 $\times10^{8}\,L_{\odot}$ with a line width of 200 km s$^{-1}$ at $\geq5\sigma$ in the velocity integrated map.}.  
However, such a source typically falls in the absolute UV magnitude range of $M_{\rm UV}$ $\sim$ $-22.0$ -- $-21.5$  mag (see e.g., Table 7 in \citealt{hashimoto2019}). This absolute UV magnitude range is $\sim$ 2--3 times brighter than the characteristic luminosity $L^{*}$ in the UV luminosity function at $z>6$ \citep[e.g.,][]{ono2018}, indicating that $\gtrsim$ 10-hour observing time is necessary to study the abundant, typical galaxies with $L^{*}$ or sub-$L^{*}$ luminosities. 
The second challenge is high spatial resolution observations towards these typical galaxies. 
Recent {\it Hubble Space Telescope} ({\it HST}) studies report that the typical effective radius ($r_{\rm e}$) in star-forming galaxies at $z>6$ is estimated to be $<$ 1 kpc ($\simeq0\farcs2$) \citep[e.g.,][]{holwerda2015,shibuya2015,bouwens2017,kawamata2018}. 
The ISM structure mostly comparable to the $r_{\rm e}$ scale could be resolved by ALMA high-resolution observations down to the $0\farcs02$ scale. 
However, this requires even longer observing times than $\gtrsim$ 10 hours estimated above just for the detection of the typical galaxies. 
The third challenge is the requirement of prior spectroscopic redshifts due to the narrow frequency coverage of ALMA (7.5-GHz coverage in a single tuning), which may cause potential biases. 
In most cases, the prior spectroscopic redshift is obtained from Ly$\alpha$ lines. 
While high-redshift galaxies with Ly$\alpha$ spectroscopic redshifts show weak \cii\ lines at a given star-formation rate (\citealt{carniani2018b,harikane2018b,harikane2020}; cf. \citealt{schaerer2020}), a recent study by \cite{smit2018} indicates that galaxies with no strong Ly$\alpha$ line may emit a strong \cii\ line. 
Because the fraction of Ly$\alpha$ emitters (LAEs; e.g., equivalent width of Ly$\alpha >$ 25 ${\rm \AA}$ ) is less than 30\% among star-forming galaxies with $M_{\rm UV} \sim -21.5$ mag at $z>6$ \citep[e.g.,][]{stark2011,treu2013,tilvi2014,debarros2017,pentericci2018,kusakabe2020}, follow-up observations of galaxies only with secure Ly$\alpha$ lines will systematically miss a majority of the representative population at $z>6$. 
An ALMA blind line survey is one possible solution, but novel \cii\ line emitters $z>6$ have not yet been identified due to the lack of sufficiently deep and large survey volumes \citep[e.g.,][]{matsuda2015,aravena2016b, yamaguchi2017, hayatsu2019, yan2020, romano2020, decarli2020}. 

In this paper, we report the blind detection of bright \cii\ 158-$\mu$m lines from strongly lensed multiple images of a sub-$L^{*}$ galaxy at $z=6.0719$ behind the massive galaxy cluster RXCJ0600$-$2007, drawn from ALMA Lensing Cluster Survey (ALCS). 
Making full use of large ancillary data sets, including {\it HST}, {\it Spitzer}, and VLT and with help of gravitational lensing magnification, we resolve the ISM structures and investigate the spatially resolved rest-frame UV-to-FIR continuum and the \cii\ line properties down to a $\simeq$ 300 pc scale. 
This is the first ALMA study to resolve the ISM properties in a representative ($\simeq$ sub-$L^{*}$) galaxy in the epoch of reionization. 

The structure of this paper is as follows. 
In Section~\ref{sec:data}, we overview the ALCS survey and the data sets in RXCJ0600$-$2007 as well as strong lensing mass models of the cluster.  
Section~\ref{sec:analysis} outlines methods of the blind line identification and optical--near infrared (NIR) properties of the two \cii\ line emitters at $z=6.07$.  
In Section~\ref{sec:results}, we report and discuss intrinsic characteristics of these two \cii\ line emitters with the correction of the lensing magnification. 
A summary of this study is presented in Section~\ref{sec:summary}. 
Throughout this paper, we assume the Chabrier initial mass function \citep{chabrier2003} and a flat universe with 
$\Omega_{\rm m} = 0.3$, 
$\Omega_\Lambda = 0.7$, 
$\sigma_8 = 0.8$, 
and $H_0 = 70$ km s$^{-1}$ Mpc$^{-1}$. 
We use magnitudes in the AB system \citep{oke1983}. 

\begin{figure*}
\includegraphics[trim=0cm 0cm 0cm 0cm, clip, angle=0,width=1.0\textwidth]{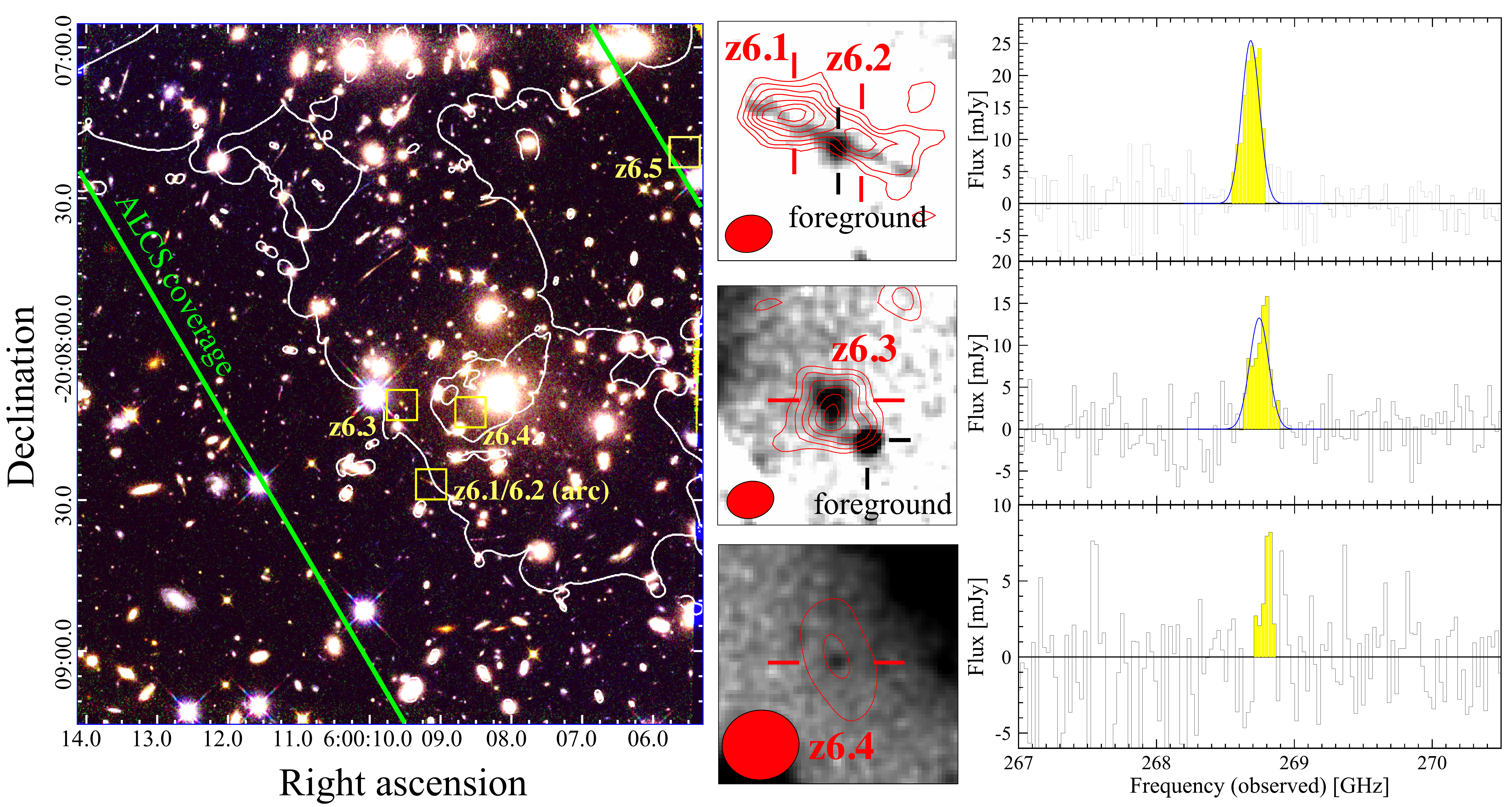}
 \caption{
{\bf Left}: 
False-color \textit{HST} image of the cluster RXCJ0600$-$2007 (red: F160W, green: F125W, blue: F814W).
The white line denotes the critical curve at $z=6.07$ estimated from our fiducial mass model. 
The green lines indicate the ALCS area coverage in this cluster, within which the relative sensitivity to the deepest part of the mosaic map is greater than 30\%.
The five multiple image positions of RXCJ0600-$z$6 are marked with the yellow  $6''\times6''$ squares. 
{\bf Middle}: 
\textit{HST}/F160W $6''\times6''$ cutouts for the multiple images of $z$6.1/6.2, $z$6.3, and $z$6.4 from top to bottom. 
The red contours denote the velocity-integrated \cii\ line intensity drawn at 1$\sigma$ intervals from $\pm2\sigma$ to $\pm$ 8$\sigma$. 
We use the natural-weighted map for $z$6.1/6.2 and $z$6.3, while we use the $uv$-tapered ($1\farcs 8\times1\farcs8 $) map for $z$6.4 to obtain the optimized S/N. 
The ALMA synthesized beams are presented in the bottom left. 
The black bars indicate foreground sources. 
The foreground source overlapping $z$6.1/6.2 is subtracted in our optical-NIR analysis with {\sc galfit} (see Section \ref{sec:opt_nir} and Appendix \ref{sec:appendix_photo}. see also N. Laporte et al. submitted). 
{\bf Right}: 
\cii\ line spectra for $z$6.1/6.2, $z$6.3, and $z$6.4 from top to bottom that we obtain in the ALCS data cube. 
The yellow shade indicates the \cii\ integration range for the velocity-integrated map whose contours are shown in the middle panel. 
The blue curve is the best-fit single Gaussian. 
\label{fig:fig1}}
\end{figure*}

\section{Data and Reduction} 
\label{sec:data}

\subsection{ALMA Lensing Cluster Survey}
\label{sec:alcs}

ALCS is a cycle-6 ALMA large program (Project ID: 2018.1.00035.L; PI: K. Kohno) 
to map a total of 88-arcmin$^{2}$ high-magnification regions in 33 massive galaxy clusters at 1.2-mm in Band 6. 
The sample is selected from the best-studied clusters drawn from {\it HST} treasury programs, i.e., the Cluster Lensing And Supernova Survey with Hubble (CLASH; \citealt{postman2012}), Hubble Frontier Fields (HFF; \citealt{lotz2017}), and the Reionization Lensing Cluster Survey (RELICS; \citealt{coe2019}).  
Observations were carried out between December 2018 and December 2019 
in compact array configurations of C43-1 and C43-2 fine tuned to recover strongly lensed (i.e., spatially elongated), low surface brightness sources. 
The 1.2-mm mapping is accomplished with a 15-GHz wide spectral scan in the ranges of 250.0--257.5 GHz and 265.0--272.5 GHz via two frequency setups to enlarge the survey volume for line-emitting galaxies. 
The spectral mode of Time Division Mode is used, which achieves the spectral resolution of $\sim$28 km s$^{-1}$ through these frequency setups. 
A full description of the survey and of its main objectives will be presented in a separate paper (in preparation).

\subsection{RXCJ0600$-$2007}
\label{sec:cluster}

RXCJ0600$-$2007 is a massive ($\sim10^{15}\, M_{\odot}$) galaxy cluster at $z=0.43$ that is included in RELICS and was
firstly identified in the Massive Cluster Survey (MACS; \citealt{ebeling2001}). 
As a part of ALCS, the ALMA observations for RXCJ0600$-$2007 were performed in January 2019, 
mapping the central area of $230''\times90''$ in 105 pointings with 46--49 12-m 
antennae 
providing baselines of 15--456 m under a precipitable water vapor (PWV) of 0.6--1.3 mm. 
J0522-3627 was observed as a flux calibrator. 
The bandpass and phase calibrations were performed with J0609-1542. 

The ALMA data were reduced and calibrated with the Common Astronomy Software Applications package version 5.4.0 ({\sc casa}; \citealt{mcmullin2007}) with the pipeline script in the standard manner. 
With the CASA task {\sc tclean}, continuum maps were produced by utilizing all spectral windows. 
The {\sc tclean} routines were executed down to the 3$\sigma$ level. 
We adopted a pixel scale of $0\farcs15$ and a common spectral channel bin of 30 km s$^{-1}$. 
The natural-weighted map achieved a synthesized beam FWHM of $1\farcs22\times0\farcs95$ with sensitivities in the continuum and the line in a 30-km s$^{-1}$ width channel of 56.9 and 932 $\mu$Jy beam$^{-1}$, respectively.  
We also produced several $uv$-tapered maps in a parameter range of $0\farcs8\times0\farcs8$ to $1\farcs8\times1\farcs8$ to obtain spatially integrated properties when necessary. 
Throughout the paper, we used the natural-weighted map unless mentioned otherwise.

\textit{HST}/ACS--WFC3 and {\it Spitzer}/IRAC observations were carried out as a part of RELICS \citep{coe2019} and {\it Spitzer}-RELICS \citep{strait2020} surveys, respectively.   {\it HST} images were obtained in the F606W (2180 s), F814W (3565 s), F105W (1411 s), F125W (711 s), F140W (736 s), and F160W (1961 s) filters.  The IRAC channel 1 ($3.6~\mu\mathrm{m}$) and channel 2 ($4.5~\mu\mathrm{m}$) integrations are approximately 10$\,$hours each.  We aligned all of the {\it HST} exposures to sources in the PanSTARRS (DR1) catalog \citep{chambers2016,flewelling2016}---which we verified is consistent with the {\it GAIA} DR2 \citep{gaia2018} astrometric frame--- and created final mosaics in a common pixel frame with 50\,mas and 100\,mas pixels for the ACS/WFC and WFC3/IR filters respectively.  We aligned the individual \textit{Spitzer} exposures to the same astrometric frame and generated final drizzled IRAC mosaics with a pixel scale of $0\farcs5$.  Further details of the {\it HST} ({\it Spitzer}) image processing with the \texttt{grizli} (\texttt{golfir}) software will be presented in Kokorev et al. (in prep).
In Figure \ref{fig:fig1}, we present the false-color {\it HST} image of RXCJ0600$-$2007.

\begin{table*}
\begin{center}
\caption{Observed FIR properties of bright \cii\ line emitters identified in RXCJ0600$-$2007}
\label{tab:line_prop}
\begin{tabular}{lcccc}
\hline 
\hline
Name          &    $z$6.1/6.2 (arc)$^{\dagger}$  &       $z$6.3     &    $z$6.4    &    $z$6.5$^{\ddagger}$       \\ \hline
R.A.     &   06:00:09.13     & 06:00:09.55        & 06:00:08.58    & 06:00:05.55         \\ 
Dec.    &   $-$20:08:26.49 & $-$20:08:11.26     & $-$20:08:12.54   & $-$20:07:20.86   \\ \hline
S/N     &  9.2               &           8.0            &    3.0               &    --                   \\
$ \nu_{\rm center}$ [GHz]         &   268.682 $\pm$ 0.011 & 268.744 $\pm$ 0.016   & (268.744)$^{\dagger\dagger}$  &    --\\
FWHM [km s$^{-1}$]& 169 $\pm$ 22       & 181 $\pm$ 34     &  (181)$^{\dagger\dagger}$    &    --\\
$ z_{\rm [CII]}$ & 6.0736$\pm$0.0003  & 6.0719$\pm$0.0004 & (6.0719)$^{\dagger\dagger}$ &    --\\
$ S_{\rm [CII]} $ [Jy km s$^{-1}$]& 4.83 $\pm$ 0.62 & 2.75 $\pm$ 0.20  & 0.44 $\pm$ 0.20&    --\\ 
$ L_{\rm [CII]}$ [$\times10^{9}\,L_{\odot}$]& 4.5 $\pm$ 0.40 & 2.3 $\pm$ 0.21   & 0.42 $\pm$ 0.19  &    --\\
$ f_{\rm 1.2 mm}$ [mJy]&   0.35 $\pm$ 0.08  & 0.20 $\pm$ 0.08 &  $<$ 0.16&    --\\
\cii\ major-axis  [$''$]   &  4.24 $\pm$ 0.82  &   1.17 $\pm$ 0.29 & --$^{\dagger\dagger}$   &    --\\
\cii\ minor-axis [$''$]   &  0.63 $\pm$ 0.51 &   0.88 $\pm$ 0.43    & --$^{\dagger\dagger}$   &    --\\
\cii\ position angle [$^\circ$]   &  71 $\pm$ 5  &   8 $\pm$ 430    & --$^{\dagger\dagger}$   &    --\\
\hline
\end{tabular}
\end{center}
\tablecomments{
{\bf S/N}: Signal-to-noise ratio at the peak pixel in the natural-weighted map, after velocity integration. The velocity integration range is denoted by the yellow shaded region in the right panel of Figure~\ref{fig:fig1}.  
{\bf  $\nu_{\rm center}$ \& FWHM}: \cii\ line peak frequency and full-width-at-half-maximum estimated from a single Gaussian fit. 
{\bf $z_{\rm [CII]}$}: Redshift of the \cii\ line emission estimated from the frequency peak.  
{\bf  $S_{\rm [CII]}$ \&  $L_{\rm [CII]}$}: The velocity integrated \cii\ line intensity and the line luminosity with optimized apertures. Here we adopt a velocity integration range of 1.5 $\times$ FWHM. 
{\bf  $f_{\rm 1.2mm}$}: Peak 1.2-mm continuum flux density in a $uv$-tapered ($2\farcs 0\times2\farcs 0$) map. We provide a 2$\sigma$ upper limit for $z$6.4. 
{\bf  \cii\ major-/minor-axis \&  position angle}: 
Deconvolved spatial size (in FWHM) and position angle of the \cii\ line in the velocity-integrated map measured with {\sc imfit}. For $z$6.1/6.2, we use a $uv$-tapered ($1\farcs 0\times1\farcs 0$) map to obtain the global scale property. 
}
$^{\dagger}$ This source is called also as RXCJ0060-arc in N. Laporte et al. (submitted). \\
$^{\dagger\dagger}$ We do not perform any profile fitting to the spectrum and the 2D spatial map of $z$6.4 due to its faintness. We adopt the FWHM and the peak frequency based on $z$6.3 for calculating the velocity-integrated intensity of the line.    \\
$^{\ddagger}$ $z$6.5 falls outside of the ALCS area coverage. 
\end{table*}

VLT/MUSE integral field spectroscopy of the RXCJ0600$-$2007 field was obtained on 26th January 2018 (ESO program ID 0100.A-0792, P.I.: A. Edge).
The 0.8-hour observation was split in three exposures of $\rm 970~seconds$ each, centered on the brightest cluster galaxy (BCG) covering $\rm 1~arcmin^2$ of the cluster core.
We use the standard MUSE reduction pipeline version 2.8.1 \citep{weilbacher2014} to create the final data-cube.
In this process, we used the self-calibration method based on the MUSE Python Data Analysis Framework \citep{bacon2016,piqueras2017} and implemented in this version of the reduction pipeline.
Finally, we applied the Zurich Atmosphere Purge \citep[ZAP,][]{kurt2016} to remove the sky residuals that were not completely removed by the MUSE pipeline.

We used the MUSE data cube to build our redshift catalog in two steps, 
similar to \citet{caminha2017, caminha2019}.
We first extracted the spectra of all sources detected in the {\it HST} imaging, and in a second step, we performed a blind search for faint-line emitters.
This procedure allowed us to measure 76	secure redshifts, of which 16 are emission from galaxies behind the cluster.
This redshift catalogue was used to identify cluster members and multiply imaged galaxies that were used in strong lens mass modeling (see Section \ref{sec:mass_model} for more details). 
In Appendix \ref{sec:appendix_spec-z}, we summarize the full spectroscopic sample from MUSE.

\section{Data analysis}
\label{sec:analysis}

\subsection{Line Identification}
\label{sec:line_prop}

We conduct a blind line search in the ALMA data cubes with the channel widths of 30 km s$^{-1}$ and 60 km s$^{-1}$. 
First, we produce three-dimensional signal-to-noise ratio (S/N) cubes by dividing each
channel with its standard deviation. 
Here we use the ALMA data cubes before the primary beam correction. 
We then search line candidates in the three-dimensional S/N cube by utilizing a python-base software of {\sc dendrogram} \citep{goodman2009} 
whose algorithm is similar to {\sc clumpfind} \citep{williams1994}. 
In {\sc dendrogram}, we obtain an initial candidate catalog of line sources that meet the following criteria: at least 10 pixels and/or channels with a pixel value of $\geq$ 2 (i.e., S/N $\geq$ 2). 
Performing the same procedure in the negative peaks in the S/N cubes under the assumption that the noise is Gaussian, 
{\sc dendrogram} evaluates the reliability of the initial line candidates based on the positive and negative properties of the peak S/N histograms, spatially integrated pixel values, and the channel width. 
This results in two reliable, bright line emitters both at $\sim$ 268.7 GHz. 
We note that these two lines are also robustly identified with an independent blind line search method of \cite{gonzalez2017c}. 
Based on morphological, redshift, and gravitational lens properties of these two line emitters obtained in detail analyses in the following subsections (Section \ref{sec:opt_nir}, \ref{sec:mass_model}, and \ref{sec:multiple}), we refer to these two line emitters as $z$6.1/6.2 (=$z6.1$ and $z$6.2) and $z$6.3 throughout this paper. 

In Figure \ref{fig:fig1}, we present the ALMA spectra and the velocity-integrated intensity (i.e., moment 0) maps of $z$6.1/6.2 and $z$6.3. $z$6.1/6.2 shows an elongated morphology with two peaks in the moment 0 map. 
Although there is a possibility that a combination of the diffuse continuum and
the noise fluctuation causes multiple peaks \citep[e.g.,][]{hodge2016}, 
we confirm in Appendix \ref{sec:appendix_sim} that the two peaks in $z$6.1/6.2 are not caused by this combination through a realistic simulation.
A single Gaussian fit to $z$6.1/6.2 and $z$6.3 in the line spectra is summarized in Table \ref{tab:line_prop}. 
Although we obtain consistent full-width-at-half-maximum (FWHM) values for the line widths between $z$6.1/6.2 and $z$6.3, their frequency peaks are slightly different by 69 $\pm$ 22 km s$^{-1}$. 
After integrating over a velocity range of 1.5 $\times$ FWHM, 
$z$6.1/6.2 and $z$6.3 have S/N values of 9.2 and 8.0 at the peak pixels, respectively. 
A single elliptical Gaussian fit over a spatial area of $6''\times 6''$ in the velocity-integrated  maps with the CASA task of {\sc imfit} yields deconvolved spatial FWHM sizes of $4\farcs 24\times 0\farcs 82$ and $1\farcs 17\times0\farcs 29$ 
for $z$6.1/6.2 and $z$6.3, respectively. 
To obtain the integrated property, here we use a $uv$-tapered ($1\farcs 0\times1\farcs 0$) map for $z$6.1/6.2 in {\sc imfit}. 
From line free channels, the continuum is also detected in the $uv$-tapered map ($2\farcs 0\times2\farcs 0$) at $4.5\sigma$ and $2.5\sigma$ level from $z$6.1/6.2 and $z$6.3, respectively.  
We also summarize the {\sc imfit} results and the continuum flux density in Table \ref{tab:line_prop}. 
Further analyses for the continuum emission are presented in N. Laporte et al. (submitted). 

\begin{table*}
\begin{center}
\caption{Observed {\it HST} and IRAC photometry of the multiple images of RXCJ0600-$z$6}
\label{tab:hst_irac_photo}
\begin{tabular}{ccccccccc}
\hline 
\hline
ID &  F606W & F814W & F105W & F125W & F140W & F160W &  3.6$\mu$m  & 4.5 $\mu$m  \\
              &  ($\mu$Jy)  &  ($\mu$Jy)  &  ($\mu$Jy)  &  ($\mu$Jy)  &  ($\mu$Jy)  &  ($\mu$Jy)  &  ($\mu$Jy)    &  ($\mu$Jy)   \\ \hline
$z$6.1/6.2 (arc) &  $<$ 0.07  &  0.32 $\pm$ 0.04 &  1.17 $\pm$ 0.07  &  1.41 $\pm$ 0.13  &  1.42 $\pm$ 0.11  &  1.34 $\pm$ 0.07  & 8.18 $\pm$ 0.42  &  6.02 $\pm$ 0.34  \\ 
$z$6.3 &   0.05 $\pm$  0.02 & 0.27 $\pm$ 0.02    &  0.99 $\pm$ 0.03  &  1.18 $\pm$ 	0.06 & 	1.17 $\pm$ 0.05 & 1.19 $\pm$ 0.03 & 5.46 $\pm$ 0.20    &	4.17 $\pm$ 0.17  \\
$z$6.4 &   $<$  0.02        &  0.07 $\pm$ 0.02    &  0.13 $\pm$	0.03 & 0.14 $\pm$	0.05 &	0.17 $\pm$	0.04 &	0.13 $\pm$ 0.03  & $<0.59$      &   $<$ 0.46$^{\dagger}$  \\
$z$6.5 &   $<$  0.02        &  0.19 $\pm$ 0.02   &  0.51 $\pm$ 0.03  &  0.63 $\pm$ 0.05 & 	0.66 $\pm$ 0.04 & 0.68 $\pm$ 0.03 & 0.40 $\pm$	0.18 & 	0.50 $\pm$ 0.15  \\
\hline
\end{tabular}
\end{center}
\tablecomments{
The photometry is performed with separate strategies for four lensed images to account for the crowded cluster field and varying degrees of extended source morphology (see Appendix \ref{sec:appendix_photo}). For non-detection, we list the upper limit at the 2$\sigma$ level.  
}
$\dagger$ We obtain 0.34 $\pm$ 0.23 $\mu$Jy which we replace the 2$\sigma$ upper limit.  
\end{table*}

\begin{figure*}
\includegraphics[trim=0cm 0cm 0cm 0cm, clip, angle=0,width=1.0\textwidth]{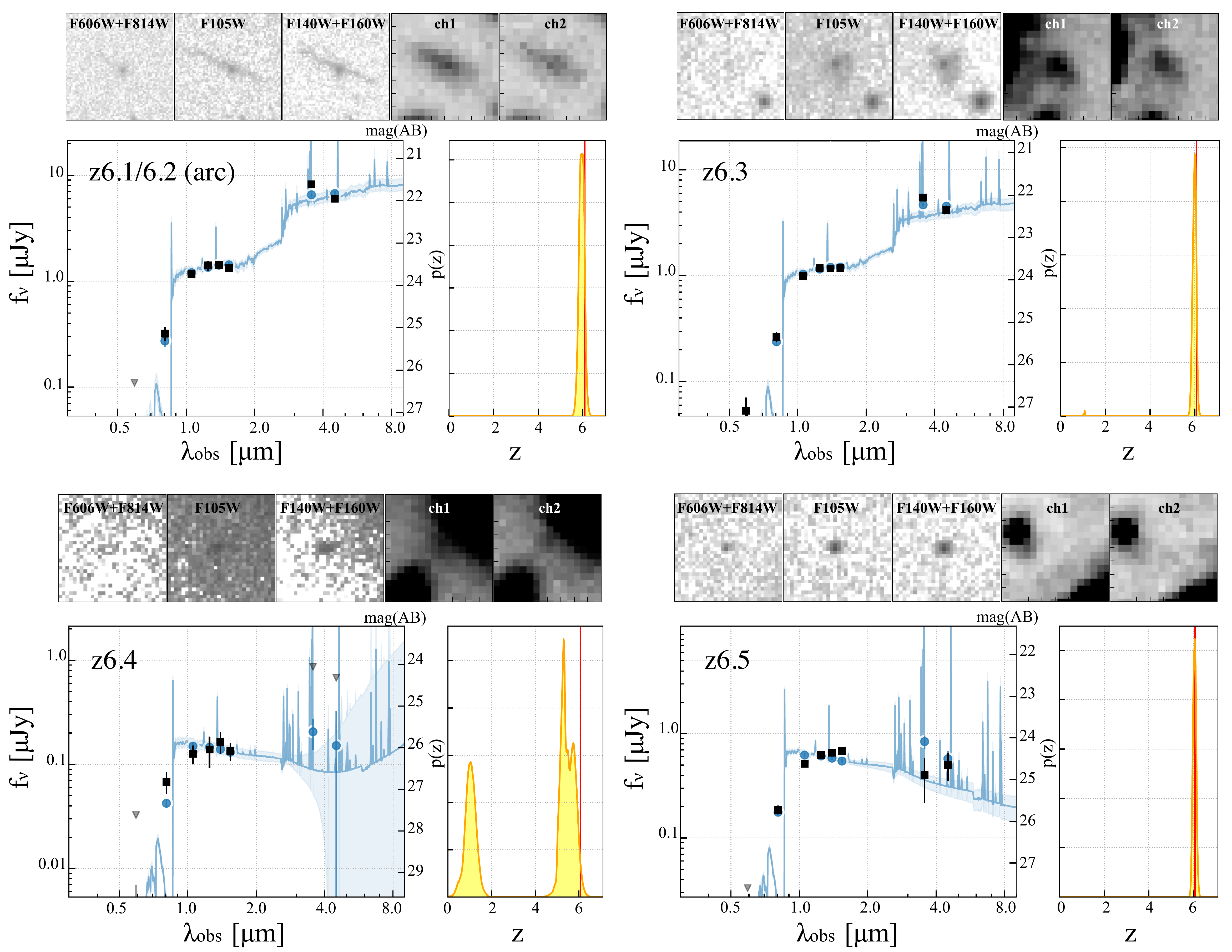}
\vspace{-0.7cm}
 \caption{
Observed optical-NIR properties of $z$6.1/6.2, $z$6.3, $z$6.4, and $z$6.5 that are predicted  as multiple images of a background LBG at $z\sim6$ consistently from different mass models (except for z6.5 whose identification as a multiple image is tentative). 
{\bf Top:} Cutouts of the {\it HST} ($3''\times3''$, except for $z$6.1/6.2 with $6''\times6''$ ) and {\it Spitzer} ($8''\times8''$) images. Some of the HST images are integrated one. 
The filter name is presented at the top. 
{\bf Bottom:} {\it HST} and {\it Spitzer} photometry (black square) and the best-fit templates, where gray triangles are the upper limits. 
The sum of individual {\sc eazy} templates is shown in the light blue curve.
The yellow shaded region is the probability distribution of the photometric redshift $p(z)$ from the SED fit. 
The red line indicates the spectroscopic redshift from the \cii\ lines at $\sim$268.7 GHz.  
A 3.6-$\mu$m excess feature in $z$6.1/6.2 and $z$6.3 is explained by the contamination of the strong \oiii$\lambda$5007 and H$\beta$  lines that are often observed in $z\sim6$ galaxies \citep[e.g.,][]{roberts-borsani2016,harikane2018b}. 
\label{fig:sed}
}
\end{figure*}

\subsection{Optical-NIR Counterparts}
\label{sec:opt_nir}

The bright lines of $z$6.1/6.2 and $z$6.3 at $\sim$268.7 GHz could be CO or \cii\ \citep[e.g.,][]{decarli2020}. 
To determine which line corresponds to $z$6.1/6.2 and $z$6.3, 
we investigate their optical to near-infrared (NIR) properties.  
In the top panel of Figure \ref{fig:sed}, we show optical-NIR {\it HST} images around $z$6.1/6.2 and $z$6.3. 
From the line peak positions, 
we identify clear counterparts in the optical-NIR images within a spatial offset of $\sim0\farcs1$ for both $z$6.1/6.2 and $z$6.3. 
Both counterparts have a noticeable dropout feature blueward of $\sim1\,\mu$m, and the one near $z$6.1/6.2 shows a highly elongated shape aligned with the elongated shape in the 268.684-GHz line. 
In the highly elongated object near $z$6.1/6.2, we identify a compact source at the center whose optical-NIR color is distinct from the other parts of the elongated object and indicative of an overlapping foreground object by chance. 
We carefully model and subtract this foreground object (see Appendix \ref{sec:appendix_photo}) to study the elongated object near $z$6.1/6.2 in the following analysis. 
In the northeast from $z$6.3, we also identify a nearby compact object.  This has a photometric redshift of 0.50$^{+0.05}_{-0.32}$ (Kokorev et al. in prep.), presumably one of the member galaxies of RXCJ0600-2007 ($z=0.43$), but does not affect the photometry of $z$6.3. 

We conduct optical-NIR photometry and spectral energy distribution (SED) analyses for these counterparts. 
We perform the aperture photometry and summarize results in Table \ref{tab:hst_irac_photo}. 
The detail procedure of the aperture photometry is described in Appendix \ref{sec:appendix_photo}. 
With the aperture photometry results, we conduct SED fitting using the {\sc eazy} code \citep{brammer2008}\footnote{http://github.com/gbrammer/eazy-py}. 
We fit the photometric flux densities and their uncertainties with linear combinations of templates derived following \cite{brammer2008} but adopting Flexible Stellar Population Synthesis models as the basis \citep{conroy2009, conroy2010}. 
We adopt the dust attenuation law of \cite{kriek2013} with $\delta=0$ (i.e., a \citealt{Calzetti2000} shape with an additional 2175 ${\rm \AA}$ dust feature). 

In Figure \ref{fig:sed}, we show probability distributions of photometric redshifts for the optical-NIR counterparts of $z$6.1/6.2 and $z$6.3. 
We find that $z$6.1/6.2 and $z$6.3 both have peak probabilities close to $z=6.0$, in excellent agreement with the bright-line detection at $\sim$ 268.7 GHz if it is the \cii\ 158 $\mu$m line at $z=6.07$. 
In this case, observed line luminosities $L_{\rm line}$ (i.e., without the correction of the lensing magnification) are estimated to be 4.5 $\pm$ 0.4 $\times10^{9}\,L_{\odot}$ and 2.3 $\pm$ 0.2 $\times10^{9}\,L_{\odot}$ for $z$6.1/6.2 and $z$6.3, respectively. 
With a standard modified blackbody at $z=6$ with a peak dust temperature $T_{\rm d}=$ 38 K \citep[e.g.,][]{faisst2020b} and a dust emissivity index $\beta_{\rm d}=1.8$ (e.g., \citealt{chapin2009,planck2011}), 
we also obtain the observed values of rest-frame FIR luminosities $L_{\rm FIR}$ to be 5.8 $\pm$ 1.3 $\times10^{11}\, L_{\odot}$ and 3.3 $\pm$ 1.3 $\times10^{11}\, L_{\odot}$ and subsequently line to rest-frame FIR luminosity ratios to be 7.8  $\times10^{-3}$ and 6.8 $\times10^{-3}$, for $z$6.1/6.2 and $z$6.3, respectively. 
These ratios are consistent with the typical range of the \cii\ line $L_{\rm [CII]}$ and $L_{\rm FIR}$ ratio ($L_{\rm [CII]}/L_{\rm FIR}$) among local galaxies \citep[e.g.,][]{brauher2008, diaz-santos2013}, which also supports the bright lines at $\sim$ 268.7 GHz being the \cii\ line.  
Based on the source redshift at $z=6.07$, we also confirm that star-formation rate (SFR) estimates  are consistent between the SED fitting with the dust attenuation correction and the summation of the rest-frame UV ($L_{\rm UV}$) and $L_{\rm FIR}$ following the work of \cite{bell2005} scaled to the Chabrier IMF, 
\begin{equation}
{\rm  SFR} \,[M_{\odot}\, {\rm yr^{-1}}] = 1.09\times10^{-10}(L_{\rm FIR}+2.2L_{\rm UV}). 
\end{equation}
Although the $z$6.3 solution shows a small non-zero probability of being at $z\sim1$, 
we also identify a 3.6-$\mu$m excess feature in both $z$6.1/6.2 and $z$6.3 which is often observed in $z\sim6$ galaxies due to the contamination of the strong \oiii$\lambda$5007 and H$\beta$  lines \citep[e.g.,][]{roberts-borsani2016,harikane2018b}. 
Therefore, the high-$z$ solution at $z\sim6$ is likely favored.  
Other possibilities for the bright line along with the high-$z$ solution might be CO(16-15) at $z=5.85$ and CO(17-16) at $z=6.28$. 
However, recent ALMA studies derive constraints on ratios of $L_{\rm CO(16-15)}/L_{\rm FIR}$ and $L_{\rm CO(17-16)}/L_{\rm FIR}$ $\lesssim$ 3$\times10^{-4}$ among luminous quasars at similar redshifts \citep{carniani2019}. 
This indicates that $L_{\rm line}/L_{\rm FIR}$ of $z$6.1/6.2 and $z$6.3 are nearly 1.5-dex higher than the typical range, strongly disfavoring the possibilities of CO(16-15) and CO(17-16) lines. 
We thus conclude that $z$6.1/6.2 and $z$6.3 are \cii\ line emitters at $z=6.07$. 
Note that we confirm that the \cii\ line solution is further supported by the lens models, intrinsic physical properties (see Section \ref{sec:multiple} and \ref{sec:int_phy}), and follow-up Gemini/GMOS spectroscopy (N. Laporte et al. submitted). 

Based on the redshift of $z=6.07$, we also examine the Ly$\alpha$ line in the MUSE data cube around $z$6.1/6.2 and $z$6.3. 
We do not identify any Ly$\alpha$ features neither around $z$6.1/6.2 nor $z$6.3. 
With the rest-frame UV luminosity, this provides 3$\sigma$ upper limits of the Ly$\alpha$ equivalent width ($EW_{\rm Ly\alpha}$) at 4.4 ${\rm \AA}$ and 3.7 ${\rm \AA}$ for $z$6.1/6.2 and $z$6.3, respectively.  
Given the dust continuum detection and the redshift, the absence of the bright Ly$\alpha$ line would be ascribed to dust and/or neutral hydrogen in interstellar and intergalactic media. This emphasizes the importance of the ALMA blind line search which enable studies of galaxies irrespective of their Ly$\alpha$ line properties in particular in the epoch of reionization.

\subsection{Mass Model}
\label{sec:mass_model}

To study intrinsic physical properties of the \cii\ line emitters $z$6.1/6.2 and $z$6.3, 
we construct several mass models for the galaxy cluster RXCJ0600--2007 ($z = 0.430$), 
using independent algorithms including {\sc glafic} \citep{oguri2010}, {\sc Lenstool} \citep{jullo2007}, and Light-Traces-Mass (LTM; \citealt{zitrin2015}). 
Multiple images are selected based on the morphology and colors of galaxies in the {\it HST} images taken with RELICS, guided by  y mass models. 
These models also exploit the MUSE spectroscopic redshift catalog (see Section \ref{sec:cluster}) for redshift information of some multiple image systems as well as secure identifications of cluster member galaxies. 
These models adopt nearly identical sets of multiple image systems for constructing the mass models and provide almost consistent predictions for multiple image positions and magnification factors. 
A brief summary of these mass models is also presented in N. Laporte et al. (submitted), while full details will be given in a separate paper (in preparation). 
In this paper, we adopt the mass model of {\sc glafic} as a fiducial model for our analyses and here describe its construction below, although we also use results of {\sc Lenstool} and LTM models to evaluate uncertainties in magnification factors.

We construct the mass model with {\sc glafic} in the same manner as in \cite{kawamata2016}. 
Our mass model consists of cluster-scale halos and cluster member galaxies. 
We place the cluster-scale halos at the positions of the three brightest cluster member galaxies in the core of the cluster. The position of one of the three cluster-scale halos is treated as a free parameter, whereas those of the other two cluster-scale halos are fixed to the galaxy positions. The cluster-scale halos are modeled by an elliptical Navarro-Frenk-White \citep[NFW; e.g.,][]{navarro1997} profile.
The cluster member galaxies are selected using both photometric redshifts of galaxies measured from {\it HST} images \citep{coe2019} as well as galaxy colors. 
The position and shapes of the member galaxies are fixed to those derived from 
the {\it HST} image and treat their velocity dispersions and truncation radii using a pseudo-Jaffe ellipsoid as model parameters assuming a scaling relation \citep[see][for more details]{kawamata2016}.
In order to achieve a good fit, a member galaxy located at (R.A., Dec.)=(06:00:10.664, $-20$:06:50.65) that produces multiple images is treated as a separate component, again assuming a pseudo-Jaffe ellipsoid.
An external shear term, which provides a modest improvement of the mass modeling result, is also included in the mass modeling of this cluster.  
After including the multiple images presented in Section~\ref{sec:multiple} that are confirmed with help of our preliminary mass models, there are  positions of 26 multiple images for eight sets of multiple image systems (five multiple image sets with spectroscopic redshifts) that we adopt as constraints.
We optimize the parameters of the mass model based on a standard $\chi^{2}$ minimization and determine the best-fit mass model assuming a positional error of $0\farcs6$ for each multiple image to account for perturbations by substructures in the cluster that are not included in our mass model, and estimate the statistical error using the Markov-chain Monte Carlo method. Our best fitting model has $\chi^2=20.0$ for 17 degree of freedom.
Interested readers are referred to \cite{kawamata2016} for more specific mass modeling procedures using {\sc glafic}.

Note that we do not include the foreground object overlapping $z$6.1/6.2 in our mass models, 
because of the absence of its spectroscopic redshift.  
We will discuss the potential contribution of the foreground object to the morphology and magnification factor of $z$6.1/6.2 in Section \ref{sec:morph}.

\subsection{Multiple Images}
\label{sec:multiple}

\begin{figure}
\includegraphics[trim=0cm 0cm 0cm 0cm, clip, angle=0,width=0.45\textwidth]{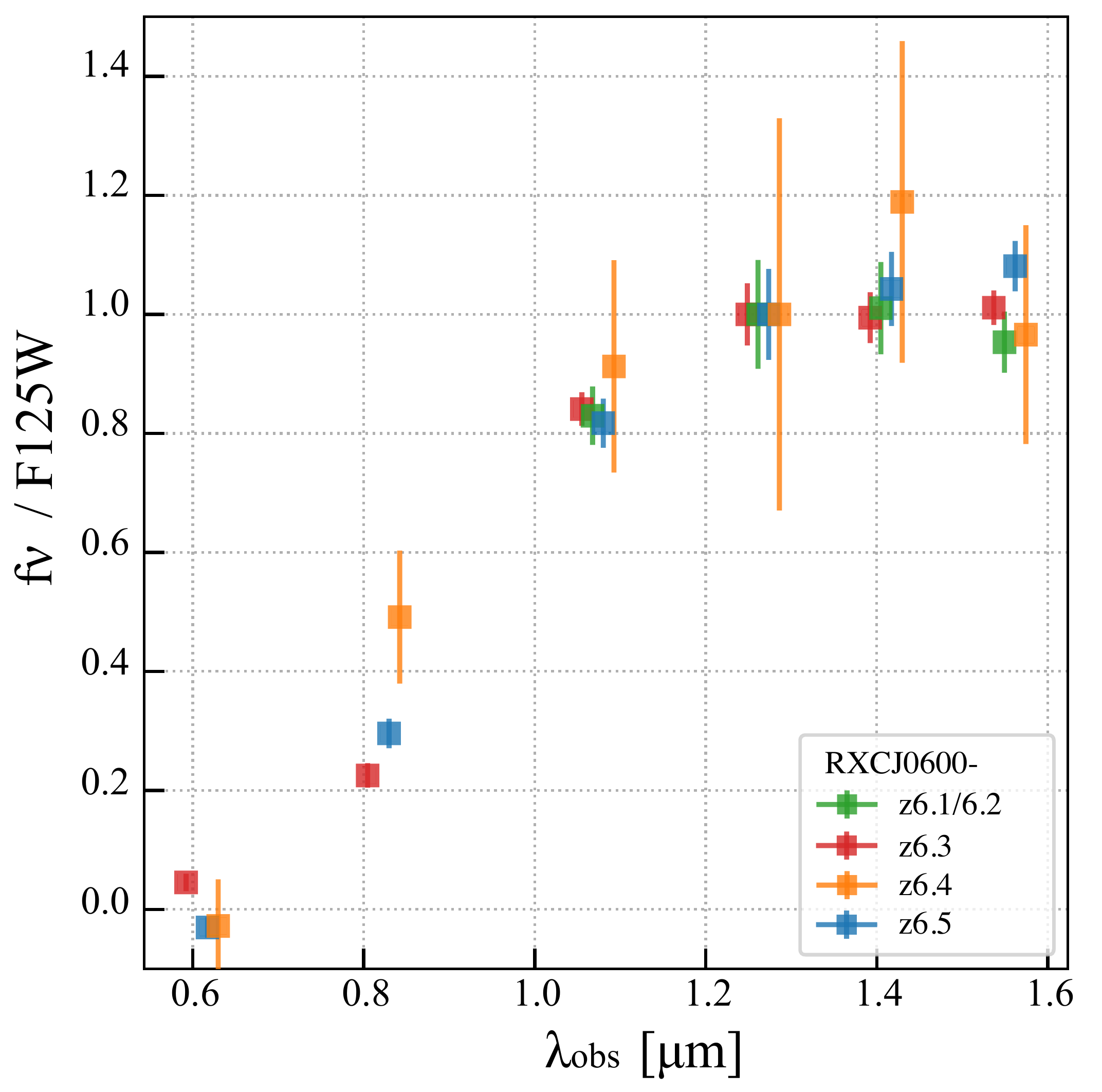}
 \caption{
Observed SEDs of the multiple images of RXCJ0600-$z$6 in the {\it HST} bands normalized by the F125W band.  The green, red, brown, and blue squares present $z$6.1/6.2, $z$6.3, $z$6.4, and $z$6.5, respectively. 
\label{fig:sed_color}}
\end{figure}

From all of our mass models, we consistently obtain the following two predictions: 
i) $z$6.1/6.2 consists of a pair of two multiple images of a galaxy at $z\sim6$ behind RXCJ0600-2007, 
and 
ii) $z$6.1/6.2 and $z$6.3 are also multiple images of the galaxy. 
The prediction of ii) is consistent with the \cii\ morphology that has two close peaks. 
In fact, we confirm in Appendix \ref{sec:appendix_z6.1-6.2} that \cii\ line spectra produced at these two peaks show line profiles consistent with each other. 
In addition, almost the same optical-NIR SED shapes between $z$6.1/6.2 and $z$6.3 in Figure \ref{fig:sed} support the prediction of ii). 
Although we identify the slight velocity shift between $z$6.1/6.2 and $z$6.3 by 69 $\pm$ 22 km s$^{-1}$ (see Section \ref{sec:line_prop}), the offset is much smaller than the typical FWHM range of the \cii\ line among $z\sim$ 4--6 galaxies evaluated in the ALPINE survey ($\sim$120--380 km s$^{-1}$; \citealt{bethermin2020}), suggesting that the slight velocity shift is explained by the differential magnification at different regions of the galaxy. 
We thus interpret $z$6.1/6.2 and $z$6.3 as multiple images of the \cii\ line emission at $z=6.07$  from a Lyman-break galaxy (LBG) behind RXCJ0600-2007.
Hereafter we refer to the background LBG as RXCJ0600-$z$6.

\begin{table*}
\begin{center}
\caption{Observed Physical Properties of the multiple images of RXCJ0600-$z$6}
\label{tab:phy_prop}
\begin{tabular}{ccccccccc}
\hline 
\hline
ID                       & $z_{\rm phot}$ & $z_{\rm spec}$  & $M_{\rm UV}$         & SFR          &           $M_{\rm star}$            &  $A_{\rm v}$  & $\mu_{\rm whole}^{\dagger}$ & $\mu_{\rm local}^{\dagger}$   \\
                           &           &           & (mag)    &  ($M_{\odot}$ yr$^{-1}$)   &  ($\times10^{9} \, M_{\odot}$)        & (mag)        &   &   \\ \hline
$z$6.1/6.2 (arc) &   5.95$^{+0.11}_{-0.13}$  & 6.0734 $\pm$ 0.0003  & $-$23.23 $\pm$ 0.07  &  135$^{+45}_{-23}$   &  41.9$^{+0.4}_{-10.1}$ &  0.07$^{+0.11}_{-0.05}$   & 29$^{+4}_{-7}$ & 163$^{+27}_{-13}$ \\ 
$z$6.3  &  5.99$^{+0.07}_{-0.09}$   &  6.0719 $\pm$ 0.0004  &  $-$23.06 $\pm$ 0.06  &   114$^{+11}_{-26}$   &  20.1$^{+1.8}_{-2.7}$  &  0.18$^{+0.09}_{-0.11}$  & 21$^{+14}_{-7}$ & --\\ 
$z$6.4    &   5.25$^{+0.48}_{-4.24}$   &  (6.0719)  &  $-$21.02 $\pm$ 0.11  &  2.6$^{+1.2}_{-0.1}$   &  0.23$^{+0.12}_{-0.01}$              &  1.80$^{+2.25}_{-1.78}$  & 3.3$^{+2.4}_{-1.2}$ & --   \\ 
$z$6.5                 &   6.05$^{+0.07}_{-0.07}$  &  --            &   $-$22.48 $\pm$ 0.05  &  8.7$^{+0.7}_{-0.1}$   &  0.77$^{+0.05}_{-0.01}$  &  0.03$^{+0.12}_{-0.01}$  & 4.2$^{+1.8}_{-1.3}$ & -- \\ 
\hline
\end{tabular}
\end{center}
\tablecomments{
Physical properties obtained from the SED fitting of the observed photometry with {\sc eazy} without correcting for the lensing magnification. 
The intrinsic physical properties after the correction of the lensing magnification are summarized in Table \ref{tab:int_prop}. 
Based on the conversion from the UV and FIR luminosity of \cite{bell2005} scaled to the Chabrier IMF, we obtain consistent SFR estimates of 188 $\pm$ 15 $M_{\odot}\,{\rm yr^{-1}}$ and 120 $\pm$ 14 $M_{\odot}\,{\rm yr^{-1}}$ for $z6.1/6.2$ and $z6.3$, respectively. Because the UV luminosity dominates in both $z6.1/6.2$ and $z6.3$, a $\pm$10 K difference in the $T_{\rm d}$ assumption for the $L_{\rm FIR}$ calculation (Section \ref{sec:opt_nir}) changes these SFR estimates by $\sim$ 0.1 dex. 
}
$^{\dagger}$ We define $\mu_{\rm whole}$ and  $\mu_{\rm local}$ as follows: \\
$\mu_{\rm whole} = $ (observed luminosity of the multiple image) / (overall luminosity of the intrinsic galaxy) \\ 
$\mu_{\rm local} = $ (observed luminosity of the multiple image) / (local luminosity of the strongly lensed, sub region near the caustic line), \\
where the sub region corresponds to the dashed rectangle area in Figure \ref{fig:lens_model}. 
The errors are evaluated from the minimum to maximum range among our independent mass models. 
\end{table*}

Subsequently, the different models also predict two additional multiple images of RXCJ0600-$z$6, 
the positions of which we present in Figure \ref{fig:fig1}, where we identify corresponding optical-NIR objects. 
We refer to these potential multiple images as $z$6.4 and $z$6.5. 
Note that different mass models predict a consistent position for $z$6.4, 
while they predict different positions for $z$6.5 with a scatter in a $\sim12''$ scale. 
In this paper, we focus an optical-NIR object as $z$6.5 predicted by one of our mass models, but it should be regarded as tentative, as we will discuss below. 

To investigate whether multi-wavelength properties of these potential multiple images are similar to $z$6.1/6.2 and $z$6.3, 
we conduct the aperture photometry for $z$6.4 and $z$6.5 in the optical-NIR bands. 
The detail procedure of the aperture photometry is again described in Appendix \ref{sec:appendix_photo}. 
In Table \ref{tab:hst_irac_photo} and  Figure \ref{fig:sed_color}, we summarize the photometry results and the optical-NIR colors normalized by the photometry at the F125W band, respectively. 
We find that $z$6.4 and $z$6.5 have similar optical-NIR SEDs with $z$6.1/6.2, $z$6.3 within the errors, consistent with our mass model predictions as multiple images at $z\sim6$. 
We further perform the optical-NIR SED fitting to $z$6.4 and $z$6.5 in the same manner as $z$6.1/6.2 and $z$6.3. 
In Figure \ref{fig:sed}, we also show the optical-NIR SED fitting results of $z$6.4 and $z$6.5.  
While the $z$6.4 photometry also allows for a $z\sim6$ solution, 
the possibility of much lower redshifts cannot be excluded due to the large uncertainties from its faint property and the potential contamination of the nearby BCG (see Appendix \ref{sec:appendix_phot_z6.4}). 
$z$6.5 has a well-localized peak probability at $z\sim6$, 
though the {\it HST}-{\it Spitzer} color of $z$6.5 is much bluer than seen for the bright images $z$6.1/6.2 and $z$6.3. In fact, an IRAC source at the location of $z$6.5 should be easily detected if it has with the same color as those of the other images.  

Because $z$6.4 falls in the ALCS area coverage, 
we also examine whether the \cii\ line emission is detected from $z$6.4 with a frequency consistent with $z$6.1/6.2 and $z$6.3. 
In Figure \ref{fig:fig1}, we also show the ALMA Band 6 spectrum of $z$6.4 based on an optimized aperture with a radius of $1\farcs 5$. 
We find that $z$6.4 has a tentative line detection (3.0$\sigma$) at the consistent frequency with $z$6.1/6.2 and $z$6.3. 
Moreover, $z$6.4 has an asymmetry line profile (the brighter peak at the higher frequency side) which is consistent with the line profile of $z$6.3. 
These results strengthen the case that $z$6.4 is indeed one of the multiple images of the background LBG. 

Based on these results, we find that the identification of $z$6.4 as one of the multiple images is relatively secure from the consistent predictions of the mass models as well as the line detection at the consistent frequency. 
On the other hand, from the different predicted positions among different mass models and the disagreements in the {\it HST}-{\it Spitzer} color with other multiple images, 
the interpretation of $z$6.5 being another multiple image is not secure and should be taken with caution until a spectroscopic redshift is obtained in follow-up observations. 
We thus use the positions of $z$6.1/6.2, $z$6.3, and $z$6.4 as constraints in deriving our best-fit mass models. 
We present the critical curve at $z=6.07$ from the best-fit mass model of {\sc glafic} in the left panel of Figure \ref{fig:fig1}. 
We summarize the \cii\ line properties and the SED fitting results for all these multiple images in Table \ref{tab:line_prop} and Table \ref{tab:phy_prop}, respectively.

\subsection{Physical Properties of RXCJ0600-$z$6}
\label{sec:int_phy}

\begin{figure*}
\includegraphics[trim=0cm 0cm 0cm 0cm, clip, angle=0,width=1.0\textwidth]{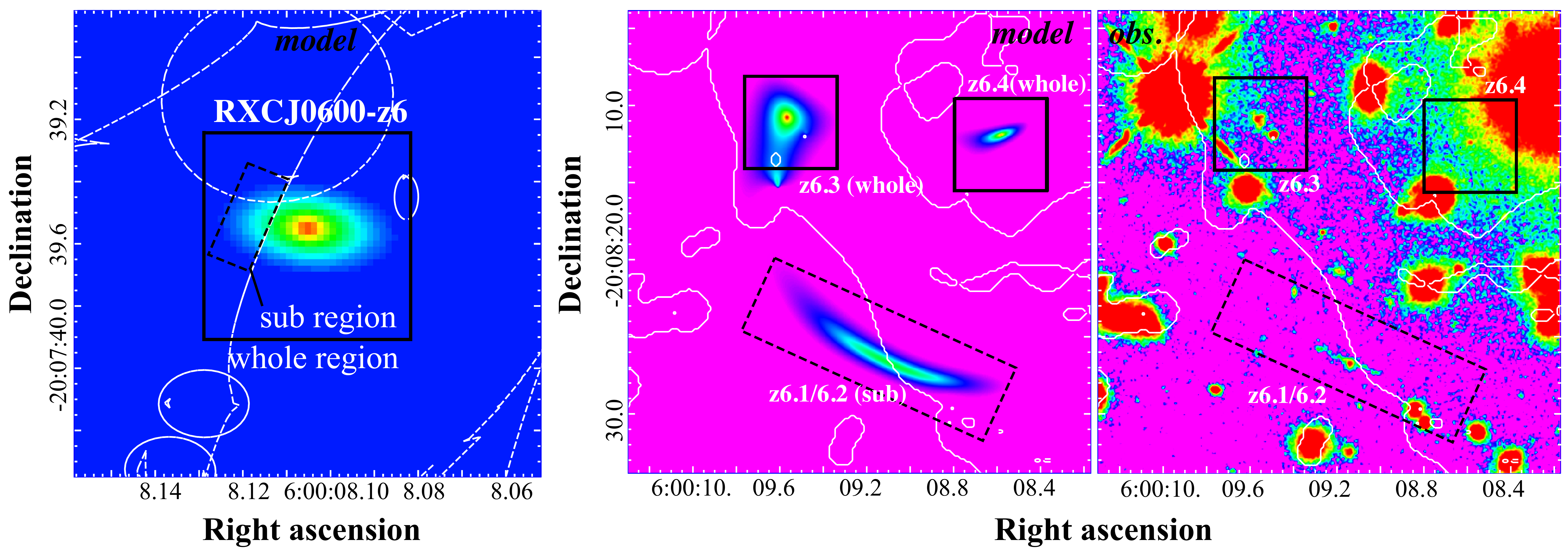}
 \caption{
{\bf Left:} 
The best-fit 2D S\'ersic profile (effective radius in major axis = 1.2 kpc, axis ratio = 0.49, S\'ersic index $n=$ 2.5) and coordinate (R.A., Decl. = 6:00:08.12, $-$20:07:39.55) of the lensed LBG RXCJ0600-$z$6 at $z=6.0719$ in the source plane based on the fiducial mass model.  
The fitting is performed based on the standard $\chi^{2}$ minimization only with the $1\farcs 7\times1\farcs 7$ {\it HST}/F160W cutout of $z$6.3. 
The white dashed curves denote the caustic lines at the source redshift. 
{\bf Middle:} 
The multiple images of RXCJ0600-$z$6 in the image plane, $z$6.1/6.2, $z$6.3, and $z$6.4, predicted by the fiducial mass model with the best-fit 2D S\'ersic profile in the left panel. 
The black boxes show the $6''\times6''$ areas around $z$6.3 and $z$6.4 at the same positions as the yellow boxes in Figure \ref{fig:fig1}. 
The dashed black rectangle denotes the $6''\times17''$ area around $z$6.1$z$6.2, 
which approximately corresponds to the dashed black rectangle shown in the left panel along the caustic line. 
The white line indicate the critical curve. 
{\bf Right:}
The {\it HST}/F160W image showing the multiple images of $z$6.1/6.2, $z$6.3, and $z$6.4. 
The image size is the same as the middle panel. 
The color and symbols are assigned in the same manner as the middle panel. 
\label{fig:lens_model}}
\end{figure*}

\begin{table}
\begin{center}
\caption{Intrinsic physical properties of strongly lensed LBG of RXCJ0600-$z$6}
\label{tab:int_prop}
\begin{tabular}{lcc}
\hline 
\hline
Name    &    \multicolumn{2}{c}{RXCJ0600-$z$6 }  \\ 
Region     &  Whole   &  Sub                      \\ 
Counter image & $z$6.3        &   $z$6.1/6.2    \\
               &     (1)                    &  (2)                         \\ \hline
RA         & 06:00:08.11          &   06:00:08.13          \\ 
Decl.     & $-$20:07:39.65       & $-$20:07:39.53        \\ 
$z_{\rm spec}$      &  6.0719$\pm$0.0004   &    6.0734$\pm$0.0003$^{\dagger}$  \\
 EW$^{\rm rest}_{Ly\alpha}$  [${\rm \AA}$]    &         $<4.4$          &    $<3.7$       \\
$L_{\rm [CII]}$ [$\times10^{8}\, L_{\odot}$]  &     1.1$^{+0.7}_{-0.5}$        &  0.3$^{+0.1}_{-0.1}$                               \\
$M_{\rm UV}$  [mag]    & $-$19.75$^{+0.55}_{-0.44}$ &  $-17.70^{+0.17}_{-0.09}$  \\
SFR [$M_{\odot}$ yr$^{-1}$] &  5.4$^{+4.5}_{-2.9}$  & 0.8$^{+0.4}_{-0.2}$     \\
$M_{\rm star}$  [$\times10^{8}\,M_{\odot}$]  &  9.6$^{+6.0}_{-4.6}$  &  2.6$^{+0.2}_{-1.0}$ \\
$A_{\rm v}$ [mag] & 0.18$^{+0.09}_{-0.11}$  & 0.07$^{+0.11}_{-0.05}$                  \\ 
$r_{\rm e}$ [kpc]    & 1.2 $^{+4.1}_{-0.1}$    &  --                     \\
$n$       & 2.5$^{+1.2}_{-0.1}$             &  --                     \\
axis ratio  & 0.49$^{+0.03}_{-0.02}$ & --                     \\
PA [$^{\circ}$]  & 84$^{+2}_{-2}$ & --                     \\
$M_{\rm dyn}$  [$\times10^{9}\,M_{\odot}$]  & 3 $\pm$ 1 & -- \\ 
$M_{\rm gas}$  [$\times10^{9}\,M_{\odot}$]  &  2 $\pm$ 1  & -- \\  
$f_{\rm gas}$  [\%]  &  $\sim$ 50--80   &  -- \\
\hline
\end{tabular}
\end{center}
\tablecomments{
(1) The physical properties related to the whole region of the galaxy that we obtain by applying $\mu_{\rm whole}$ (Table \ref{tab:phy_prop}) to the observed properties of $z$6.3. The best-fit S\'ersic profile ($r_{\rm e}$, $n$, and axis ratio) is not estimated by using $\mu_{\rm whole}$, but by optimizing the intrinsic 2D surface brightness profile in the source plane to match the 2D surface brightness profile of $z$6.3 in the image plane with {\sc glafic} (see Section \ref{sec:int_phy}). 
The circularized effective radius is estimated to be 0.8$^{+2.9}_{-0.1}$ kpc, which is consistent within $\sim$1--2$\sigma$ errors with an independent 2D surface brightness profile fit on the image plane in N. Laporte et al. (submitted).
We calculate $M_{\rm gas}$ by subtracting $M_{\rm star}$ from $M_{\rm dyn}$, which is consistent with another estimate from the empirical calibration with the \cii\ luminosity \citep{zanella2018} of $(3\pm1)\times10^{9}\,M_{\odot}$ (see Section \ref{sec:kin}). 
(2) The physical properties related to the local scale of the galaxy in the sub region near the caustic line, by applying $\mu_{\rm local}$ (Table \ref{tab:phy_prop}) to the observed properties of $z$6.1/6.2. \\
$\dagger$ The sub region is red-shifted by 69 $\pm$ 22 km s$^{-1}$, which agrees well with the velocity gradient identified in the whole scale of RXCJ0600-$z$6 (see Section \ref{sec:kin}). 
}
\end{table}

\begin{figure}
\includegraphics[trim=0cm 0cm 0cm 0cm, clip, angle=0,width=0.45\textwidth]{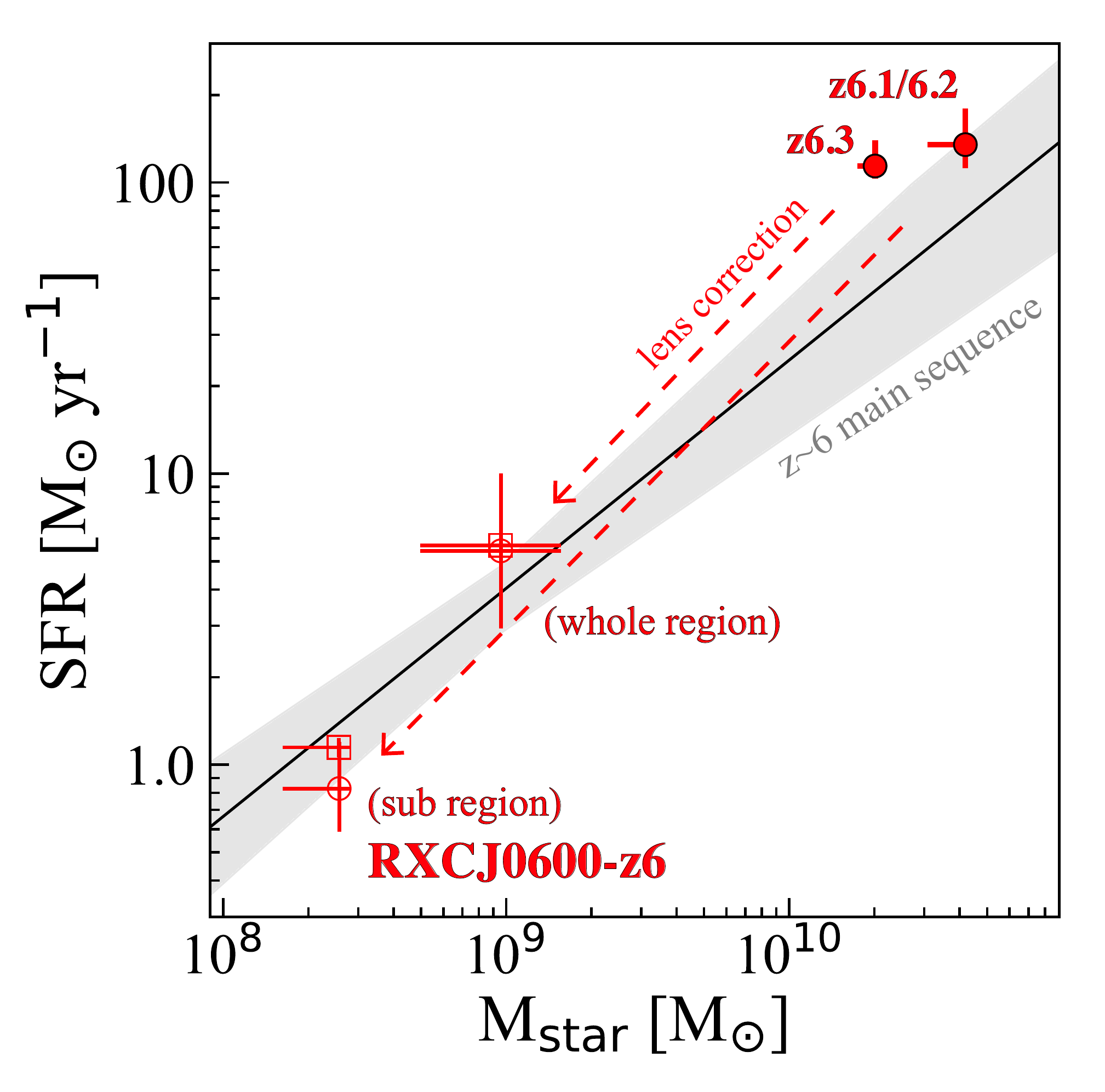}
 \caption{
SFR--$M_{\rm star}$ relation.  
The red filled and open circles indicate the relations before and after applying the correction of the lensing magnification to the SED fitting results, respectively, for $z$6.1/6.2 and $z$6.3.  
The errors include the uncertainty from the mass models (Tabel \ref{tab:phy_prop}). 
The red open squares indicate SFR estimates after the lensing magnification correction based on the conversion from the UV and FIR luminosity of \cite{bell2005} scaled to the Chabrier IMF.
The black line and the gray shaded region denote the best-fit relation for $z\sim6$ galaxies and its $1\sigma$ uncertainty evaluated in \cite{iyer2018}. 
\label{fig:sfr-ms}}
\end{figure}

The configuration of the multiple images is helpful to obtain the precise information about the source position and its surface brightness profile in the source plane. 
Here we estimate the intrinsic two-dimensional (2D) surface brightness profile by fitting the {\it HST} images assuming the fiducial mass model. 
Specifically, we first produce a $1\farcs 7\times1\farcs 7$ cutout {\it HST}/F160W image of $z$6.3. 
With a single S\'ersic profile model in the source plane, we then obtain the best-fit effective radius $r_{\rm e}$ = 1.2$^{+4.1}_{-0.1}$ kpc (major axis), axis ratio of 0.49$^{+0.03}_{-0.02}$, position angle = 84$^{+2\circ}_{-2}$, S\'ersic index $n$ = 2.5$^{+1.2}_{-0.1}$, and central coordinate of (RA, Decl.)=(6:00:08.12, $-$20:07:39.55) based on standard $\chi^{2}$ minimization. 
Because we find a degeneracy between $r_{\rm e}$ and $n$, 
here we restrict the S\'ersic index to the range of $1<n<4$ in the fitting. 
We do not use $z$6.1/6.2 and $z$6.4 for the fitting due to the complicated morphology and the contamination of the diffuse emission from the nearby BCG, respectively. 
We note that here we ignore the clumpy structure of $z$6.3 for the moment, which we will discuss later. 
We list these best-fit S\'ersic profile results in Table \ref{tab:int_prop}.

In Figure \ref{fig:lens_model}, we show the best-fit 2D S\'ersic profile in the source plane and its multiple images in the image plane. We find that a single S\'ersic profile 
well reproduces not only $z$6.3, but also $z$6.4 and $z$6.1/6.2 whose elongated shape is interpreted as a result of the source crossing the caustic line in the source plane and stretched over $\sim6''$ scale in the image plane \citep[e.g.,][]{vanzella2020}.
This interpretation is consistent with the slight difference in the line peak frequencies and the line profiles between $z$6.1/6.2 and $z$6.3 (Section \ref{sec:line_prop}), because the sub region of the galaxy can have different kinematics properties compared to the whole galaxy. 
By calculating the ratio of the spatial areas between the source and image planes, the magnification factors for $z$6.1/6.2 and $z$6.3 in our fiducial model (average of the three independent models) are estimated to be $\sim$150 (163) and $\sim$35 (21), respectively. 
The observed luminosity of $z$6.1/6.2 is 33 (29) times brighter than the intrinsic overall luminosity of RXCJ0600-$z$6 due to the strong gravitational lensing effect near the caustic line. By comparing  physical properties of $z$6.3 and $z$6.4 that are both tracing the whole region of the lensed galaxy, we confirm that our independent mass models agree in the ratio of magnification factors between $z$6.3 and $z$6.4 in the range of 6.1--6.7, which is consistent with the observed $L_{\rm [CII]}$ ratio of 5.7 $\pm$ 2.7 between $z$6.3 and $z$6.4. 
These results validates our best-fit mass models and 2D S\'ersic profile in the source plane. 

To be conservative, we adopt the average value of the magnification factors 
and evaluate its uncertainty from the minimum to maximum values among our independent mass models, when we estimate the intrinsic physical properties of RXCJ0600-$z$6 in this paper. 
We list the average magnification factor and its uncertainty in Table \ref{tab:phy_prop}. 
Applying the average magnification factors to the FIR (Section \ref{sec:line_prop}) and optical-NIR (Section \ref{sec:opt_nir}) properties, 
we summarize the intrinsic physical properties in whole and sub regions of RXCJ0600-$z$6 in Table \ref{tab:int_prop}. 
Remarkably, we obtain the intrinsic absolute rest-frame UV magnitude of $M_{\rm UV}= -19.75^{+0.55}_{-0.45}$, which is $\sim3$ times fainter than $L^{*}$ of the LBG luminosity function at $z=6$ ($M_{\rm UV}=20.91^{+0.07}_{-0.06}$; \citealt{ono2018}). 
In Figure \ref{fig:sfr-ms}, we show the SFR and $M_{\rm star}$ relation of RXCJ0600-$z$6. For comparison, we also present the average relation among $z\sim6$ galaxies estimated in \cite{iyer2018} (gray shaded region). 
We find that RXCJ0600-$z$6 falls on the average relation from the sub to whole regions. 
We also find that the relation between the \cii\ line width and luminosity in RXCJ0600-$z$6 agrees with the average value among $z\sim6$ galaxies and the theoretical prediction (see Figure 10 in \citealt{kohandel2019}). 
The circularized effective radius ($r_{\rm e, circ} \equiv r_{\rm e} \times \sqrt{\rm axis\, ratio}$) of 0.84 kpc also falls in a typical range among $z\sim6$ galaxies with the similar UV luminosity (see e.g., Figure 9 of \citealt{kawamata2018}). 
These results indicate that RXCJ0600-$z$6 is an abundant, representative sub-$L^{*}$ galaxy at this epoch. 
We note that these intrinsic physical properties are consistent with independent estimates in N. Laporte et al. (submitted) within the errors, even though the SED fitting strategies are different due to the different scopes in the paper.

\begin{figure*}
\includegraphics[trim=0.1cm 0.1cm 0.1cm 0.1cm, clip, angle=0,width=1.0\textwidth]{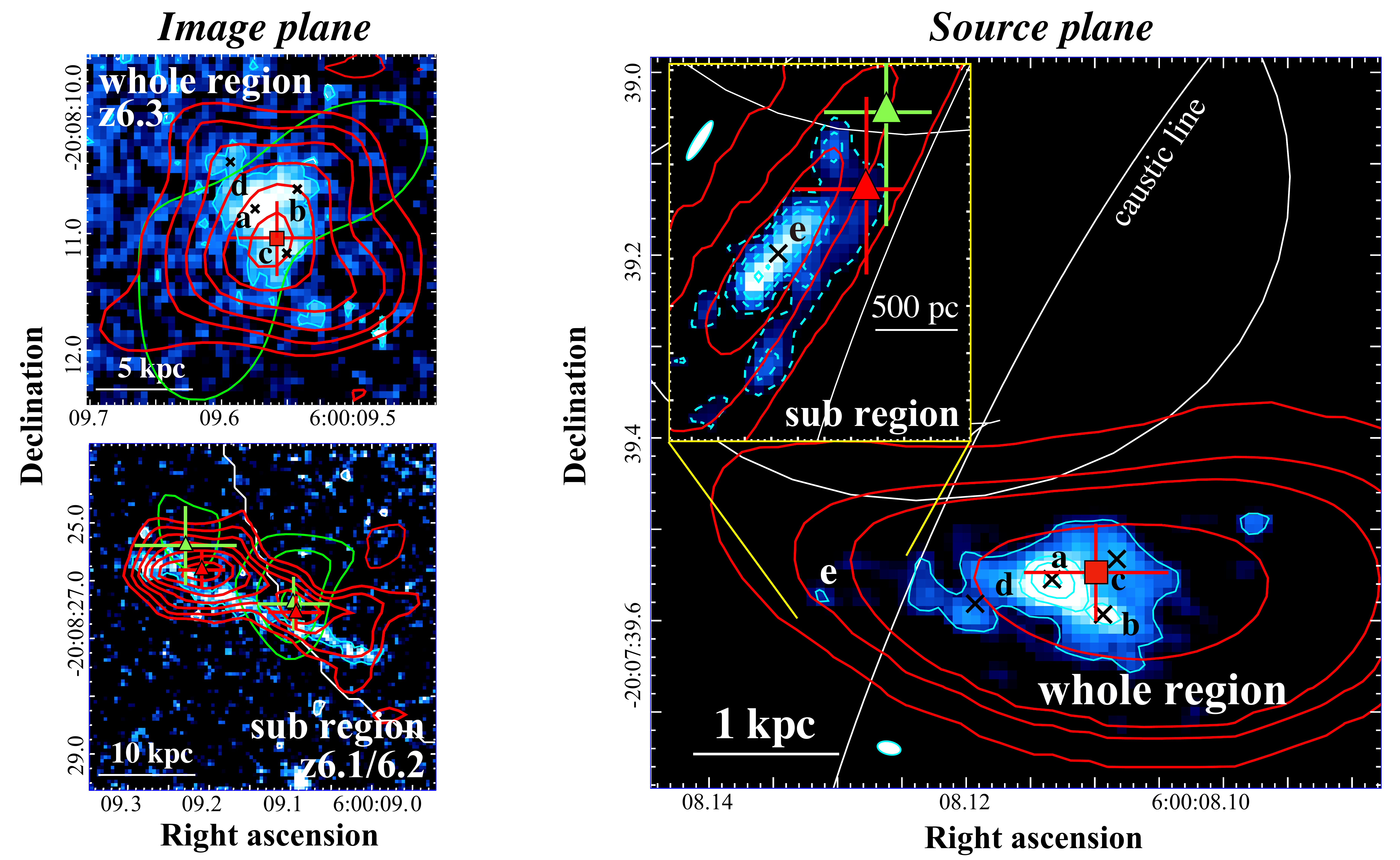}
 \caption{
{\bf Left:} 
$3''\times3''$ and $6''\times6''$ {\it HST}/F160W cutouts of $z$6.3 (top) and $z$6.1/6.2 (bottom) in the image plane. 
The small black crosses denote bright clumps (marked $a$, $b$, and $c$) and an elongated structure towards north east ($d$) in the rest-frame UV continuum. 
The dashed black cross indicates the peak position of the rest-frame UV continuum after smoothing the {\it HST} map to match the resolution with ALMA.  
The red and green contours represent the \cii\ line and the rest-frame FIR continuum from ALMA drawn at $1\sigma$ intervals from $2\sigma$ to $8\sigma$. 
The red and green squares (triangles) in $z$6.3 ($z$6.1/6.2) show the emission peak pixel positions of the \cii\ line and rest-frame FIR continuum with the 1$\sigma$ error bars, respectively.  
The cyan contours is drawn at 2$\sigma$ for the rest-frame UV continuum. 
The foreground galaxies are removed  with {\sc galfit} for the source plane reconstruction. 
Here we use the natural-weighted map for the \cii\ line, while the $uv$-tapered maps with $2\farcs0\times2\farcs 0$ and $0\farcs 8\times0\farcs 8$  for the rest-frame FIR continuum of $z6.3$ and $z6.1/6.2$, respectively. 
{\bf Right:}
 Source plane reconstruction of the \cii\ line and rest-frame UV and FIR continuum of $z6.3$. 
The color and symbols follow the same assignment as the left panel, where the cyan and red contours show 10\%, 30\%, 50\%, and 80\% of the peak. To match the spatial resolutions of ALMA and HST, the red contours are drawn from the source plane reconstruction of the de-convolved \cii\ spatial distribution obtained with {\sc imfit} (Section \ref{sec:line_prop}) that is smoothed with the HST PSF. 
The inset panel displays the source plane reconstruction of $z6.1$/6.2. The cyan (red) contours show 1\%, 3\%, 5\%, and 10\% (10\%, 30\%, and 50\%) of the peak of the whole galaxy. The luminosity-weighted center is marked with label $e$ which corresponds to the faint rest-frame UV clump at the western part of the whole galaxy. 
The white ellipses show the typical shape of the HST PSF reconstructed in the source plane. 
The error bars of the red square, triangle, and green triangle incorporate the average lensing magnification corrections and their uncertainties. 
 Note that two peaks (= triangles) in $z$6.1/6.2 in the image plane to be the multiple pair (Section \ref{sec:mass_model}), which thus correspond to one peak (= triangle) in the source plane.
\label{fig:reconstruction}}
\end{figure*}

\section{\cii\ Views from ISM to Cosmic Scales}
\label{sec:results} 

The uniquely and strongly lensed galaxy near the caustic line (Section~\ref{sec:mass_model}) allows us to study ISM properties from internal to whole scales of the host galaxy. 
For an example of the whole view of the galaxy based on $z$6.3, the spatial resolutions of the HST map of $\sim0\farcs2$ translate into $0\farcs04$ (corresponding to $\sim$250 pc at $z=6.07$) after the correction of the lensing magnification, providing sub-kpc scale ISM views. 
At the same time, the blind aspect of the ALCS survey also allows us to statistically evaluate the number density of the \cii\ line emitters at $z\sim6$ in a cosmic scale based on our successful identification of RXCJ0600-$z$6.  
In conjunction with the rest-frame UV and FIR continuum properties, we examine the \cii\ line properties from the ISM to cosmic scales and discuss whether there is common property or a large diversity among these multiple scales. 

\subsection{Spatial Distributions of UV, FIR, and \cii\ down to Sub-kpc Scale}
\label{sec:morph}

Making full use of the gravitational lensing, we investigate spatial distributions of the \cii\ line, rest-frame UV and FIR continuum on the source plane and compare them.  
In the context of similar studies so far at $z\sim$ 2--4 for bright dusty, starburst galaxies \citep[e.g.,][]{swinbank2010,swinbank2015, dye2015, spilker2016, rybak2015, tamura2015, hatsukade2015a, rybak2020, rizzo2020} and less massive galaxies \citep[e.g.,][]{mirka2017,mirka2019}, this is a first observation to resolve the ISM structure down to the sub-kpc scale for the sub-$L^{*}$ galaxy in the epoch of reionization. 

In the left panel of Figure \ref{fig:reconstruction}, we present the rest-frame UV continuum maps for $z$6.3 (i.e., whole region) and $z$6.1/6.2 (i.e., sub region) taken in the {\it HST}/F160W band with the \cii\ line (red contour) and the rest-frame FIR continuum (green contour) taken by ALMA. 
The emission peaks of the \cii\ line and rest-frame FIR continuum are marked with the red and green squares (triangles) for $z$6.3 ($z$6.1/6.2) with the 1$\sigma$ error bars\footnote{The error is estimated by the approximate positional accuracy of the ALMA map $\Delta p$ in milliarcsec, given by $\Delta p$ = 70000/$(\nu * B * \sigma$), where $\sigma$ is the peak SNR in the map, $\nu$ is the observing frequency in GHz, and $B$ is the maximum baseline length in kilometers (see Section 10.5.2 in cycle 7 ALMA technical handbook)}, respectively. 
Here we do not examine the rest-frame FIR continuum peak from $z$6.3 due to the poor significance at the $2.5\sigma$ level (see Section \ref{sec:line_prop}). 
In $z$6.3, the rest-frame UV continuum shows a clumpy structure, and thus we mark these clumps with black crosses labeled with $a$, $b$, and $c$, from brightest to faintest. 
The rest-frame UV continuum of $z$6.3 also shows an elongated structure toward the north east, which we mark with an additional black cross and label $d$. 
In $z$6.1/6.2, possible clumps are more evident in the \cii\ line and the rest-frame FIR continuum with the two-peak morphology. 
If RXCJ0600-$z$6 consists of a smooth disk, the morphology of $z$6.1/6.2 would be a single smooth-arc shape as shown in the middle panel of Figure \ref{fig:lens_model}. 
Therefore, the two-peak morphology of $z$6.1/6.2 may imply that the ISM of RXCJ0600-$z$6 near the caustic line has a clumpy structure in the source plane.
An alternative possibility is that an intrinsically smooth disk is stretched into the two-peak morphology in the image plane due to the perturbation by the foreground object overlapping $z$6.1/6.2 which is not included in our fiducial mass model. 
To check this possibility, we include the foreground object in our mass model assuming that it is a member galaxy of the cluster (see Appendix \ref{sec:appendix_phot_z6.1}) and find that the two-peak structure can indeed be reproduced, if the mass associated with the foreground object is comparable or larger than that expected from the scaling relation of the luminosity and mass for cluster member galaxies constrained in our mass modeling. 
We conclude that we need more follow-up data including the spectroscopic redshift of the foreground object in order to discriminate these two possibilities. 
We however confirm that both magnification factors of $\mu_{\rm local}$ and $\mu_{\rm whole}$ for $z$6.1/6.2 are affected only by $\sim$1--2 \% even if we include the foreground object in the mass model as one of the cluster members or outside of the cluster up to $z\sim4$. 
The other foreground object near $z6.3$ is classified as one of the member galaxies of the cluster (Section \ref{sec:opt_nir}) and predicted to produce the critical curve at the southern east part of $z6.3$ in our mass model (see the middle panel of Figure \ref{fig:lens_model}). However, we find that the \cii\ morphology at the corresponding area is not disturbed at all, suggesting that its lensing effect is negligible for $z6.3$. We thus remove this foreground object from the mass model and the HST map with {\sc galfit} in the source plane reconstruction of $z6.3$. 

In the right panel of Figure \ref{fig:reconstruction}, we present the source plane reconstruction of $z$6.3 (whole region), 
where the inset panel displays the source plane reconstruction of $z$6.1/6.2 (sub region). 
To match the spatial resolution between HST and ALMA, we create a \cii\ map from the de-convolved \cii\ spatial distribution (Section \ref{sec:line_prop}), smooth it with the point spread function (PSF) of the HST F160W band, and use this PSF-matched map for the source plane reconstruction of the \cii\ line. 
In the right panel, the white ellipses indicate the source plane reconstruction of the HST PSF whose FWHM is decreased down to $\sim200\times100$ pc and $\sim300\times 60$ pc around $z6.3$ and $z6.1/6.2$, respectively. 
The other color and symbols follow the same assignment as the left panel, where we apply the lens correction also to the error bars. 
The error bar of the \cii\ line peak position in $z$6.3 (red square) is decreased down to $\sim$300 pc. 
These results indicate that we are able to map the ISM view down to a few hundred parsec scale.
Note that our independent mass models consistently suggest that the two-peak morphology of $z$6.1/6.2 in the image plane are the pair of multiple images (Section \ref{sec:mass_model}) regardless of whether there exists the foreground object or not, 
which thus correspond to one peak in the source plane. 
We confirm that the entire morphology in both whole and sub regions of the galaxy and the emission peak positions 
does not change beyond the errors in the source plane whether we include or not the foreground galaxy overlapping $z$6.1/6.2 in the mass model as one of the member galaxies of RXCJ0600-2007. 

Firstly from the reconstruction of $z$6.3, we find on the scale of the galaxy that \cii\ line peak shows an offset of $\simeq$ 300 pc from the brightest rest-frame UV clump of a, but they are consistent at the 1$\sigma$ error level. 
With the axis ratio of 0.49 (Table \ref{tab:int_prop}),  non-parametric measurements directly on the surface brightness distributions in the source plane provide $r_{\rm e} = $ 1.1 kpc and 2.6 kpc for the rest-frame UV continuum and the \cii\ line emission, respectively, showing the spatially extended \cii\ gas structure by a factor of $\sim$ 2.4. 
These results are consistent with the recent ALMA results of \cite{fujimoto2020b} for 23 individual normal star-forming galaxies at $z\sim4$--6, whereby generally the \cii\ line is spatially more extended than the rest-frame UV continuum by factors of $\sim2$--3 without a spatial offset beyond $\simeq$ a 1-kpc scale. 
The $r_{\rm e}$ value for the rest-frame UV continuum is also consistent with the S\'ersic profile fitting results of 1.2 kpc in the source plane presented in Section~\ref{sec:int_phy}. 

Secondly from the reconstruction of $z$6.1/6.2, 
we find in the sub region of the galaxy that the \cii\ line is co-spatial with the rest-frame UV continuum again, which is 
separated by $\sim1.6$ kpc from the peak of the \cii\ line and rest-frame UV continuum from the whole region of the galaxy. 
We mark the luminosity-weighted center of the sub region with the black cross labeled $e$. 
Remarkably, we find, in the independent rest-frame UV continuum map reconstructed from $z6.3$, that the faint clump exists exactly at the position of $e$ whose peak flux density is also consistent. These agreements in the properties of the clump $e$ also support the robustness of our best-fit mass models. 
We also find that the \cii\ and rest-frame FIR peaks observed in the image plane are reconstructed in the source plane with a $\simeq$ 1 kpc offset from the luminosity-weighted center of $e$. This indicates that the faint diffuse emission or further faint clump near the caustic line is strongly lensed and more prominently visible in the image plane than the clump $e$. 
Given that \targbg\ is quantified with $r_{\rm e}=1.2$ kpc in the rest-frame UV continuum (Section \ref{sec:int_phy}), these results indicate that we are witnessing very faint \cii\ and rest-frame FIR emitting region(s) near the caustic line beyond the effective radius of the galaxy that is almost invisible in other multiple images. 
Because of the poor significance level of the rest-frame FIR continuum in $z6.3$, we cannot conclude whether the rest-frame FIR continuum detected in $z6.1/6.2$ corresponds to the outskirt emission of the whole galaxy or the localized emission at the sub region of the galaxy. 

Interestingly, the brightest peaks  of the \cii\ line and the rest-frame FIR continuum in $z$6.1/6.2 appear on opposite sides in the image plane (Left bottom panel of Figure \ref{fig:reconstruction}). 
In $z$6.1/6.2, the magnification factor is generally the same on either side. 
The clear difference identified in the \cii\ line strength at the high significance levels ($8.2\sigma$ and $5.4\sigma$) suggests the existence of substructure of the mass distribution along the line-of-sight of $z$6.1/6.2, which is so-called {\it flux-ratio anomaly} \citep[e.g.,][]{mao1998}. 
This is consistent with our interpretation that the central compact object in the optical-NIR bands in $z$6.1/6.2 is the foreground object which is responsible for this {\it flux-ratio anomaly}. 
However, if the \cii\ and rest-frame FIR emitting regions are identical in the source plane, the flux ratio should be the same between the \cii\ and rest-frame FIR emission in the image plane. Although the current error bars of the spatial positions are large, this independent observable of the flux ratio suggests that the faint \cii\ and rest-frame FIR emitting regions are physically offset in the sub region of RXCJ0600-$z$6. This potential separation and the detailed ISM structure in RXCJ0600-$z$6 must be addressed in future deeper and higher-resolution observations. 

\begin{figure*}
\includegraphics[trim=0cm 0cm 0cm 0cm, clip, angle=0,width=1.\textwidth]{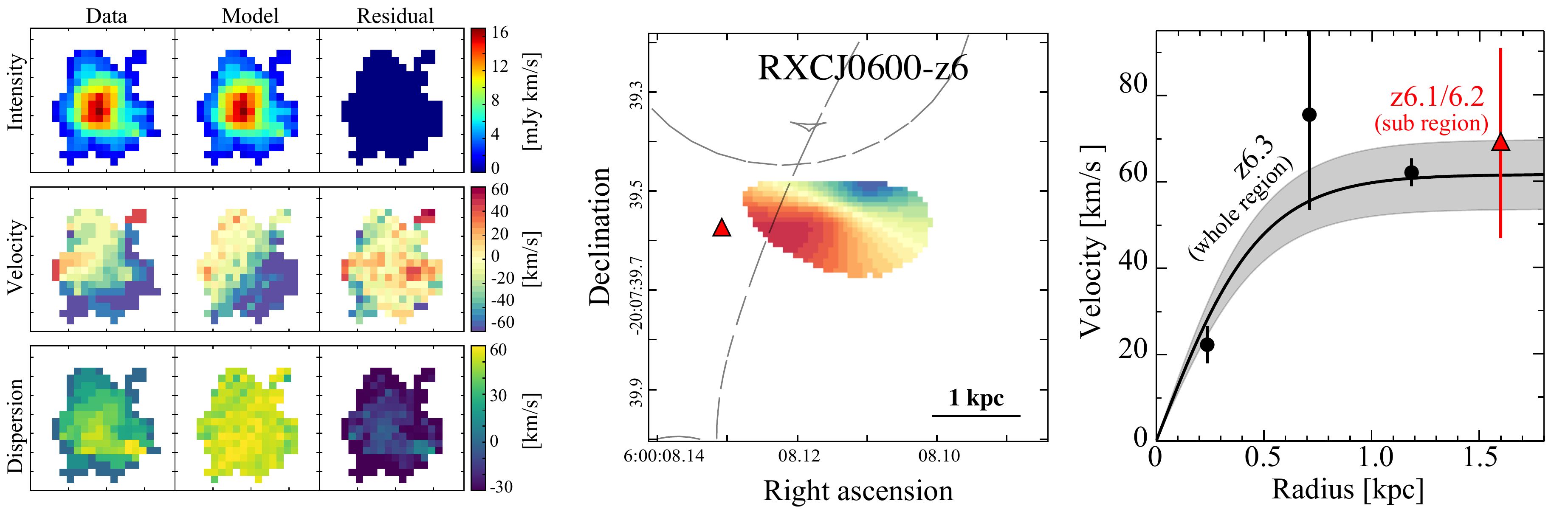}
 \caption{
Kinematic properties of \targbg. 
{\bf Left:}
Velocity-integrated (i.e., intensity; top), velocity-weighted (middle), and velocity-dispersion (bottom) maps of $z6.3$ are shown in the left column. The best-fit 3D model with $^{\rm 3D}${\sc barolo} and the residual maps are presented in the middle and right columns, respectively. 
The image size is $3''\times3''$. 
{\bf Middle:} 
Source-plane reconstruction of the intrinsic velocity-weighted map of $z$6.3 that was obtained with {\sc galpak3D}. 
We limit the reconstruction up to the radius of $1\farcs4$  from the \cii\ intensity peak in the image plane.  
The grey dashed line is the caustic line. 
The red triangle shows the luminosity-weighted center of the sub region of the galaxy. 
{\bf Right:}
Observed radial velocity profile of $z6.3$ extracted from the three annuli defined for the $^{\rm 3D}${\sc barolo} analysis (black circles) and the spatial and velocity offsets of the sub region of the galaxy observed in $z6.1/6.2$ (red triangle). 
The black line shows the best-fit, together with the associate $1\sigma$ error (grey shade), of the radial velocity profile of $z6.3$ obtained from {\sc galpak3d}. 
\label{fig:kin}}
\end{figure*}

\subsection{Kinematics via \cii }
\label{sec:kin}
We also examine the kinematics of RXCJ0600-$z$6 via the bright \cii\ line emission. Here we focus on the \cii\ kinematics of $z6.3$ to characterize the gas kinematics of the whole galaxy. 
In the left panel of Figure \ref{fig:kin}, we present the velocity-integrated (top), velocity-weighted (middle), and velocity-dispersion (bottom) maps of $z6.3$ in the image plane. We evaluate the root-mean-square noise level from the data cube and create these maps with a three-dimensional (3D) mask of all signal above the 2$\sigma$ level. 
We find that the \cii\ line has the velocity gradient from $-$100 to +45 km s$^{-1}$ in east south to west north with its intensity extended up to a radius of $\sim1\farcs4$. 
Assuming the line width estimate of $z6.3$ (Table \ref{tab:line_prop}) and a potential error of $\sim$30 km~s$^{-1}$ for the velocity gradient due to the spectral resolution of our ALMA data cube (Section \ref{sec:alcs}), we obtain $\Delta v_{\rm obs}/2\sigma_{\rm tot}=0.94\pm0.26$, where $\Delta v_{\rm obs}$ and $\sigma_{\rm tot}$ are the full observed velocity gradient  (uncorrected for inclination) and the spatially-integrated velocity dispersion, respectively. 
With an approximate diagnostic for the classification of rotation-dominated and dispersion-dominated systems, $\Delta v_{\rm obs}/2\sigma_{\rm tot}=0.4$ \citep{schreiber2009}, we find that $z6.3$ is classified as the rotation-dominated system. Note that the beam smearing effect generally makes the velocity gradient [dispersion] underestimated [overestimated] in spatially low-resolution maps (see e.g., Figure 7 of \citealt{diteodoro2015}). This strengthens the argument that $z6.3$ is the rotation-dominated system from the increased $\Delta v_{\rm obs}/2\sigma_{\rm tot}$ value without the beam smearing effect.

To study the rotation kinematics,  
we analyze our data in the image plane with softwares of $^{\rm 3D}${\sc barolo} \citep{diteodoro2015} and {\sc galpak3d} \citep{bouche2015} that are tools for fitting 3D models to emission-line data cubes.
In the left panel of  Figure \ref{fig:kin}, we also show the best-fit 3D model and residual maps with $^{\rm 3D}${\sc barolo} by assuming three annuli for its tilted ring fitting algorithm.
We find an excellent agreement on the intensity map and that the residual velocities in velocity-weighted and velocity-dispersion maps are generally less than the spectral resolution of our ALMA data cube ($\sim$28 km s$^{-1}$; Section \ref{sec:alcs}). 
Although the residual in the velocity dispersion is relatively large near the edge of the mask, this is likely because the faint outskirt emission near the edge is masked in some velocity channels and the observed velocity dispersion is underestimated.  These results suggest that the \cii\ kinematics of $z6.3$ is well reproduced by the best-fit 3D model.  
We confirm that an independent 3D modeling of a single exponential disk with {\sc galpak3d} also provides the best-fit values of the rotation velocity, the velocity dispersion, and the inclination fully consistent within errors with the $^{\rm 3D}${\sc barolo} results. 
We summarize the details for the 3D modeling and the results in Appendix \ref{sec:appendix_kin}.

In the middle panel, we present the velocity-weighted map of $z6.3$ in the source plane via the reconstruction in the same manner as Section \ref{sec:morph}.  To understand the intrinsic picture without the beam smearing effect, here we use the best-fit intrinsic (i.e., resolution free) map obtained from {\sc galpak3d} for the reconstruction. In the right panel, we also show the \cii\ radial velocity extracted from the three annuli with $^{\rm 3D}${\sc barolo} (black circle) as well as the best-fit (black line) and the 1$\sigma$ error (gray shade) of the rotation curve in the tanh formalization obtained from {\sc galpak3d}. The correction the lensing magnification is applied to the radius scale. For comparison, the spatial and velocity offsets of $z6.1/6.2$ are shown with the red triangle in both middle and right panels. We find that $z$6.1/6.2 agrees with the velocity gradient of $z$6.3 within the errors, which is consistent with our interpretation that $z$6.1/6.2 is the sub region of RXCJ0600-$z6$. This suggests that the clump $e$ in the sub region of the galaxy (Section \ref{sec:morph}) is likely a small star-forming region within the rotation disk of the host galaxy.

For the rotation-dominated system, 
we obtain the dynamical mass $M_{\rm dyn}$ of  $(3\pm1) \times10^{9}\,M_{\odot}$ based on an assumption of the disk-like gas potential distribution, following the equation (4) in \cite{mirka2020} 
\begin{eqnarray}
\left (\frac{M_{\rm dyn}}{M_{\odot}} \right) &=& 1.16\times10^{5} \left(\frac{v_{\rm rot}}{\rm km\,s^{-1}}\right)^{2} \left(\frac{r_{\rm e}}{\rm kpc} \right),  
\end{eqnarray}
where $v_{\rm rot}$ is the rotation velocity of the gaseous disk after the inclination correction. We calculate the inclination from the axis ratio of the best-fit surface brightness profile results for the rest-frame UV continuum (Section \ref{sec:int_phy}), assuming that the higher-resolution map provides a better constrain for the inclination. We adopt $r_{\rm e}$ and $v_{\rm rot}$ from the source plane reconstruction of the \cii\ line (Section \ref{sec:morph}) and the {\sc galpak3d} results, respectively. 
We caution that the uncertainty of the inclination could remain by $\sim30\%$ even in the spatially resolved analysis \citep[e.g.,][]{rizzo2020}, and thus the uncertainty in the above $M_{\rm dyn}$ estimate could be even larger. 
Given the negligible contribution of the dark matter halo in the galactic scale, 
we estimate the molecular gas mass $M_{\rm gas}$ to be $\sim$ 1--2 $\times10^{9}\,M_{\odot}$ by subtracting $M_{\rm star}$ (Section \ref{sec:int_phy}) from $M_{\rm dyn.}$. 
It is worth noting that this $M_{\rm gas}$ range agrees with another estimate based on an empirically calibrated method in \cite{zanella2018}, given by 
\begin{eqnarray}
\left (\frac{L_{\rm [CII]}}{L_{\odot}} \right) = 10^{-1.28\,(\pm0.21)} \times \left (\frac{M_{\rm gas}}{M_{\odot}} \right)^{0.98\,(\pm0.02)}, 
\end{eqnarray}
which suggests $M_{\rm gas}= 3^{+2}_{-1} \times 10^{9}\,M_{\odot}$, despite the potentially large uncertainty of the inclination.  
These results indicate that RXCJ0600-$z6$ is a gas-rich galaxy with a high gas fraction of $f_{\rm gas}$ ($\equiv M_{\rm gas}/(M_{\rm star}+M_{\rm gas}$)) $\sim$ 50--80 \%. This is consistent with recent ALPINE results that \cii-detected ALPINE galaxies with $M_{\rm star}\sim1\times10^{9}\, M_{\odot}$ have $f_{\rm gas}$ $\sim$ 60--90 \% (see Figure 8 in \citealt{mirka2020}). 
These $M_{\rm dyn}$, $M_{\rm gas}$, and $f_{\rm gas}$ estimates are also listed in Table \ref{tab:int_prop}. 

Note that we cannot rule out the possibility that the velocity gradient is originally caused by complex dynamics with interacting, merging galaxies. 
Future higher resolution observations will confirm the smooth rotation of the disk or break the complex dynamics into the multiple components.

\begin{figure*}
\begin{center}
\includegraphics[trim=0cm 0cm 0cm 0cm, clip, angle=0,width=1.0\textwidth]{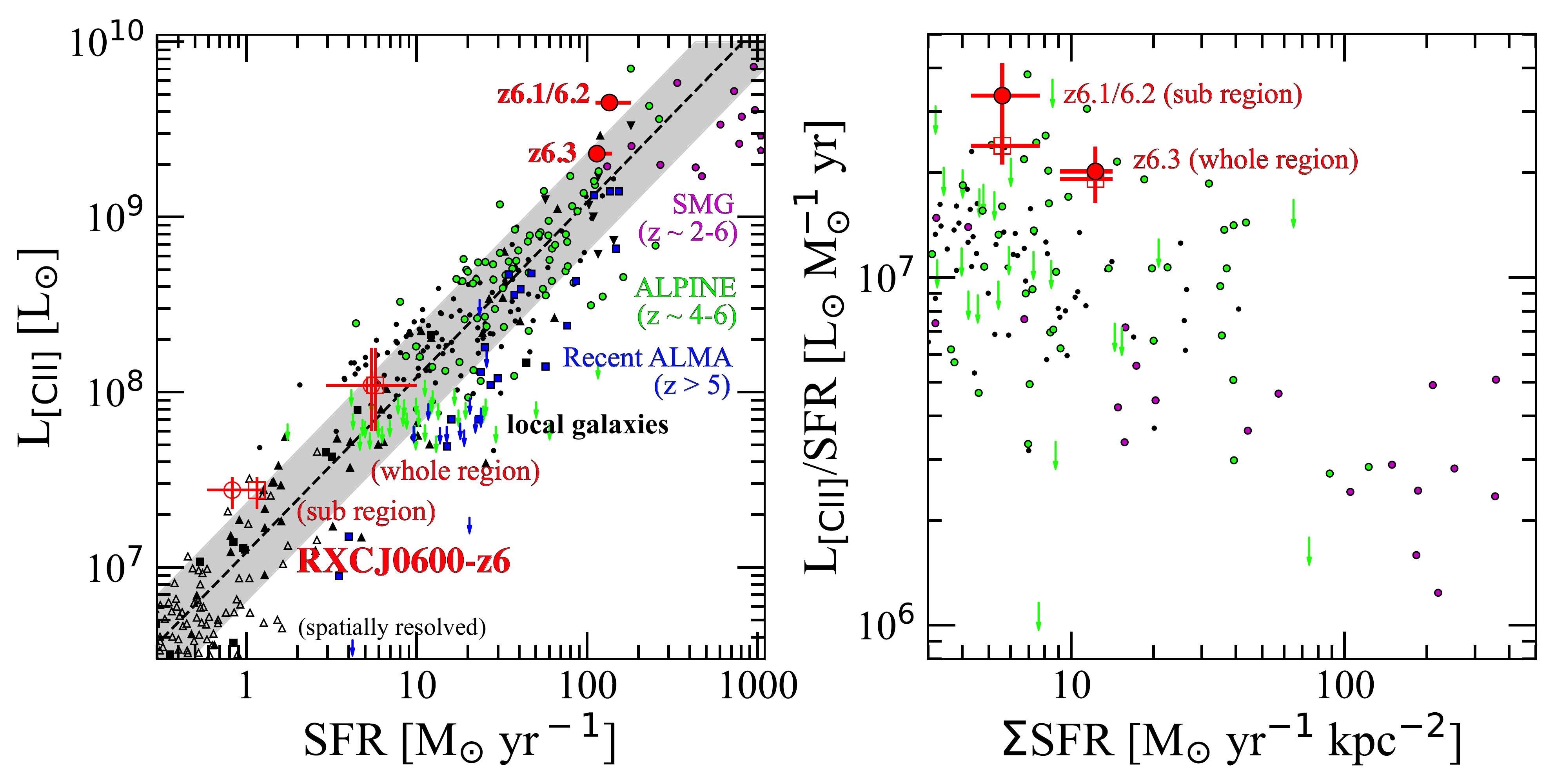}
\end{center}
\vspace{-0.2cm}
\caption{
{\bf Left:} $L_{\rm [CII]}$--SFR relation. 
The red filled and open circles indicate $z$6.1/6.2 and $z$6.3 before and after the correction of the lensing magnification, respectively. 
The errors include the uncertainty from the mass models (Tabel \ref{tab:phy_prop}). 
The red open squares indicate the SFR estimates after the correction of the lensing magnification based on the conversion from the UV and FIR luminosity of \cite{bell2005} scaled to the Chabrier IMF. 
The previous results from local to high-$z$ star-forming galaxies are shown with 
black circles (local LIRGs; \citealt{diaz-santos2013}), black squares (local dwarfs; \citealt{delooze2014}), black triangles (local spirals; \citealt{malhotra2001}), black inverse triangles ($z\sim0.3$ (U)LIRGs; \citealt{magdis2014}), green circles ($z\sim4$--6 star-forming galaxies from ALPINE; \citealt{schaerer2020}), magenta pentagons ($z\sim5$ submillimeter galaxies (SMGs); \citealt{cooke2018}), magenta circles ($z\sim$ 2--6 lensed SMGs; \citealt{spilker2016}), and blue squares (compilation of recent ALMA results for $z>5$ star-forming galaxies; \citealt{matthee2020} and \citealt{harikane2020}). 
We adopt the average relation of the low-$z$ H{\sc ii}-galaxy/starburst sample from \cite{delooze2014}, which is adjusted to the Chabrier IMF by reducing the SFR by a factor of 1.06 in the same manner as \cite{schaerer2020}. 
The arrow indicates the 3$\sigma$ upper limit. 
The $L_{\rm FIR}$ value in the literature is firstly converted into a total IR luminosity $L_{\rm TIR}$ (8--1000 $\mu$m), and then we calculate SFR by using the calibration of \cite{murphy2011}. 
The spatially resolved results ($\Sigma$SFR and $\Sigma L_{\rm [CII]}$) for local galaxies are also presented with open triangles \citep{herrera-camus2015} by assuming the area of 1 kpc$^{2}$. 
The dashed line and gray shade denote the $L_{\rm [CII]}$--SFR relation obtained from local star-forming galaxies \citep{delooze2014} and its dispersion, respectively. 
{\bf Right:} $L_{\rm [CII]}$/SFR--$\Sigma$SFR relation. 
The color assignments on the symbols are the same as the left panel.  
We define the star-forming area by a circular area of the rest-frame UV emission with a radius of $\sqrt{2}\times r_{\rm e, circ}$ for the $\Sigma$ SFR estimate. We use the factor of $\sqrt{2}$ in accordance to \cite{spilker2016}. 
For $z$6.1/6.2 and $z$6.3, we evaluate the rest-frame UV size by reducing the \cii\ size measurements with {\sc imfit} (Table \ref{tab:line_prop}) by a factor of 2 (Section \ref{sec:morph}). 
For the ALPINE sources, we use the rest-frame UV size measurement results in \cite{fujimoto2020b}. 
For local LIRGs and lensed SMGs, we use the rest-frame FIR size measurement results in \cite{spilker2016} by assuming that the star-forming activity is dominated in the rest-frame FIR emitting regions in these objects.   
\label{fig:sfr-cii}}
\end{figure*}

\subsection{SFR and $L_{\rm [CII]}$ Relation}
\label{sec:sfr-cii}

In the left panel of Figure \ref{fig:sfr-cii}, 
we show the relation between SFR and $L_{\rm [CII]}$ for $z$6.1/6.2 and $z$6.3. 
For comparison, we also show local and high-redshift galaxy results taken from the literature \citep{malhotra2001, diaz-santos2013, magdis2014, delooze2014, herrera-camus2015, spilker2016, cooke2018, harikane2020, matthee2019, schaerer2020} and the $L_{\rm [CII]}$--SFR relation obtained from local star-forming galaxies in \cite{delooze2014}. 
The observed $L_{\rm [CII]}$ of both $z$6.3 and $z$6.1/6.2 fall on the most luminous $L_{\rm [CII]}$ regime among typical (e.g., SFR $\lesssim$ 100 $M_{\odot}\,{\rm yr}^{-1}$)  high-$z$ star-forming galaxies, demonstrating the power of the gravitational lensing. 
After the correction of the lensing magnification, we find that both $z$6.3 and $z$6.1/6.2 fall slightly above, but still likely follow the SFR--$L_{\rm [CII]}$ relation of the local galaxies within the dispersion. 
This is consistent with recent ALMA results that the average SFR--$L_{\rm [CII]}$ relation among high-redshift star-forming galaxies at $z\sim4$--9 is well within the intrinsic dispersion of the local relation \citep{carniani2018b, carniani2020,schaerer2020}. 
Given that RXCJ0600-$z$6 is consistent with being an abundant, sub-$L^{*}$ galaxy at $z=6$ (see Section \ref{sec:int_phy}), these results may suggest that the SFR--$L_{\rm [CII]}$ relation, defined by local galaxies, holds from the 
spatially resolved sub-kpc ISM to the whole scales in abundant galaxies even up to the epoch of reionization. 

To further study the $L_{\rm [CII]}$--SFR relation, 
the right panel of Figure \ref{fig:sfr-cii} presents $L_{\rm [CII]}$/SFR and SFR surface density ($\Sigma$SFR). 
This relation or another relation between $L_{\rm [CII]}/L_{\rm FIR}$ and $L_{\rm FIR}$ surface density ($\Sigma L_{\rm FIR}$) 
are known to have tight anti-correlations where the deficit of the \cii\ line is explained by the high ionization state in the ISM around regions with high $\Sigma$SFR or  $\Sigma L_{\rm FIR}$ \citep[e.g.,][]{diaz-santos2013,spilker2016,gullberg2018,ferrara2019}.
Importantly, these relations are not affected by the lensing magnification, 
because the same magnification factor applies to all these values. 
We find that $z$6.1/6.2 shows a higher $L_{\rm [CII]}$/SFR ratio, 
while both $z$6.1/6.2 and $z$6.3 are consistent with the trend of the anti-correlation. 
This indicates that the difference of $\Sigma$SFR causes the difference of the \cii\ line luminosity at a given SFR between $z$6.1/6.2 and $z$6.3, 
which is likely consistent with the source plane reconstruction results in Section \ref{sec:morph}: 
the faint \cii-emitting region of $z$6.1/6.2 is separated from the bright rest-frame UV clumps by $\sim$1.6 kpc in the source plane, where the ionization state of the local ISM is thought to be moderate. 

We note that recent ALMA observations show non-detection results of the \cii\ line from similarly star-forming galaxies at $z\sim4$--9 at the same time (see upper limits in Figure \ref{fig:sfr-cii}), indicative of the existence of galaxies whose $L_{\rm [CII]}$--SFR relations are different from those of the local galaxies. 
Given the requirement of prior spectroscopic redshift with the Ly$\alpha$ line in most cases (Section \ref{sec:opt_nir}), 
those non-detections might be related to recent reports of the potential anti-correlation between $L_{\rm [CII]}$/SFR and EW$_{\rm Ly\alpha}$ (\citealt{harikane2018b,harikane2020,carniani2018b}). 
In contrast to the most cases, \targbg\ is identified in the blind survey and its physical properties (rest-frame EW$_{\rm Ly\alpha}$ $<$ 4.4 ${\rm \AA}$ )  agrees with the potential anti-correlation reported in \cite{harikane2018b}. 
Another lensed galaxy at $z=6.15$ \citep{calura2021} also follows the similar trend with relatively large rest-frame EW$_{\rm Ly\alpha}$ (60 $\pm$ 8 ${\rm \AA}$) and small $L_{\rm [CII]}$/SFR ($\sim2\times10^{5}$). 
A caution still remains that \cite{schaerer2020} report a weak dependence of $L_{\rm [CII]}$/SFR of EW$_{\rm Ly\alpha}$. 
Since the \cii\ line emissivity depends on the ISM properties such as the ionization state, metallicity, and gas density \citep[e.g.,][]{vallini2015}, the different $L_{\rm [CII]}$--SFR relations could be alternatively explained by a larger dispersion of the ISM properties in high-$z$ galaxies than in local galaxies.  
The uncertainties of the SFR estimates might contribute to the large dispersion in high-$z$ galaxies due to assumptions of the star-formation history, the dust-attenuation curve, and the stellar population age as discussed in \cite{carniani2020} and \cite{schaerer2020}. 
Another recent reports of the extended \cii\ line morphology up to a radius of $\sim$10 kpc \citep[e.g.,][]{fujimoto2019,fujimoto2020b,ginolfi2020,novak2020} might be also related to some of those non-detections, because the surface brightness of the extended emission is significantly decreased in relatively high-resolution maps \citep{carniani2020}. Based on the visibility-based stacking, the secondary extended component up to the $10$-kpc scale is estimated to have the average contribution to the total line luminosity of  $\sim$ 50 \% around star-forming galaxies \citep{fujimoto2019} and $\sim20\%$ around quasars \citep{novak2020} at $z\sim6$. 
These non-negligible contributions could matter if the request sensitivity is close to the detection limit around the 5$\sigma$ level. However, this is not the case if the carbon in the extended \cii\ gas is ionized by such as the gravitational energy in the cold stream, the shock heating in the outflow and/or inflow gas, and the AGN feedback, instead of the photoionization powered from the star-forming regions (see e.g., Section 5 of \citealt{fujimoto2019}).

\subsection{\cii\ Luminosity Function}
\label{sec:cii_LF}

\begin{figure}
\includegraphics[trim=0cm 0cm 0cm 0cm, clip, angle=0,width=0.49\textwidth]{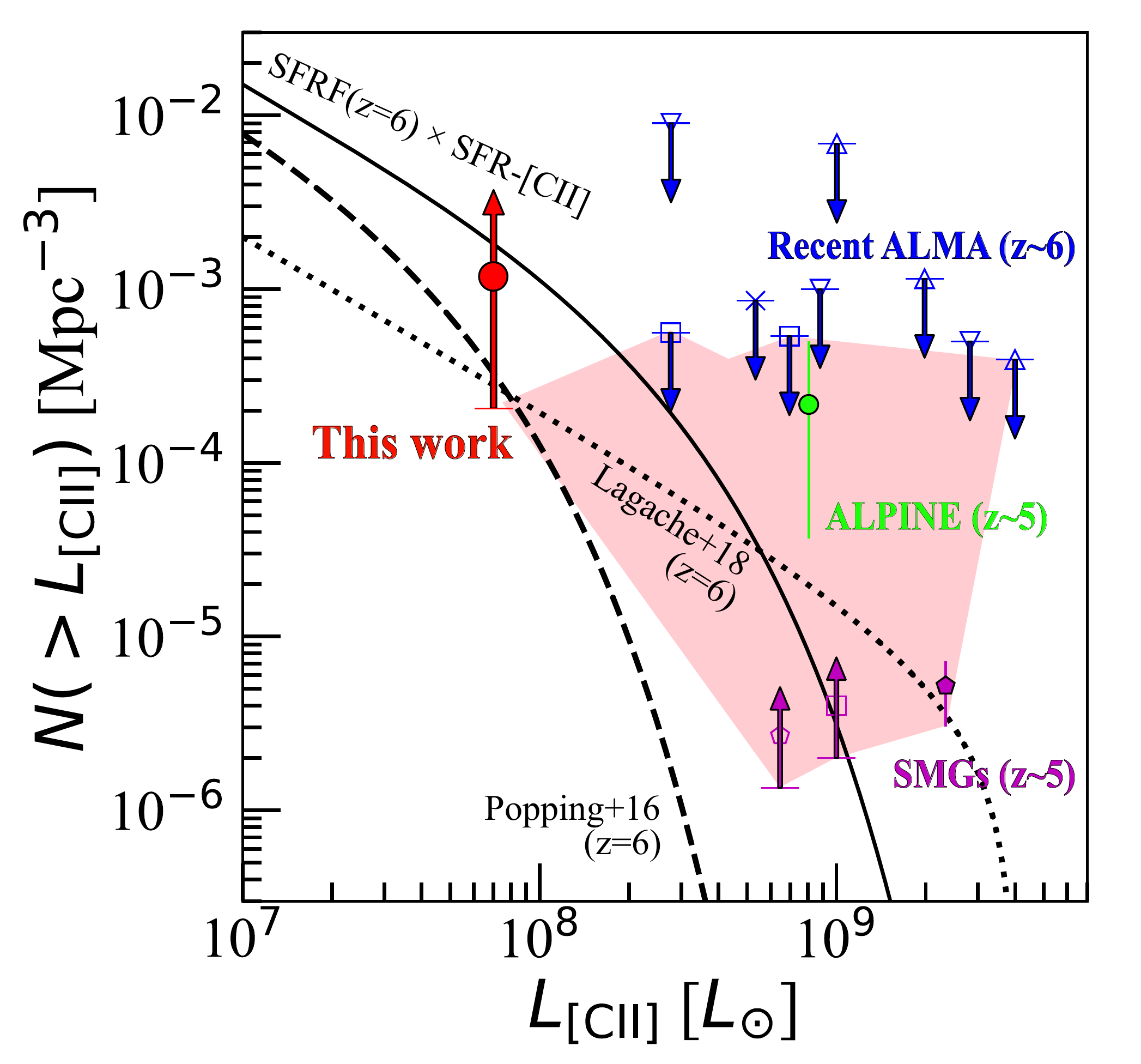}
 \caption{
Cumulative \cii\ luminosity function at $z=6$ with recent \cii\ line studies at $z>4$ . 
The red circle shows the number density of the \cii\ line emitter based on our successful detection from the strongly lensed LBG with the effective survey volume of the full ALCS data cubes composed of 33 galaxy clusters. 
The lower limit is estimated from the Poisson uncertainty at the single-sided confidence level of 84.13\% presented in \cite{gehrels1986}. 
Recent ALMA blind line survey results are presented with 
blue triangle (243 archival data cubes; \citealt{matsuda2015}), blue inverse triangle (four massive galaxy clusters; \citealt{yamaguchi2017}), blue square (ASPECS; \citealt{decarli2020}), and blue cross (SSA22; \citealt{hayatsu2017, hayatsu2019}). 
The green circle presents the ALPINE results \citep{loiacono2020,yan2020}. Here we show only the estimate from the serendipitous \cii\ line detection at $z\sim5$ whose redshift is sufficiently separated from (i.e., not associated with) the central ALPINE targets. 
The magenta square and pentagon show the serendipitous \cii\ line detection from bright SMGs at $z\sim5$ reported in \cite{swinbank2012} and \cite{cooke2018}, respectively. 
The red shade indicates the current constraints based on our and previous results so far obtained. 
For comparison, we also present 
semi-analytical model results \citep{popping2016,lagache2018} and SFR function \citep[SFRF; ][]{smit2016}, including the dust correction based on the SMC extinction law, whose SFR value is converted into $L_{\rm [CII]}$  with the local \cii--SFR relation \citep{delooze2014}. 
\label{fig:cii_LF}}
\end{figure}

A key goal of ALCS is to constrain the number density of the line emitters. Although the complete blind line survey results with all 33 fields will be presented in  a separate paper (in preparation), 
we can evaluate a lower limit of the \cii\ luminosity function at $z\sim6$ with our \cii\ line detection from the strongly lensed LBG at $z=6.0719$. 

To do this, we first measure the effective survey area using mass models for all 33 ALCS clusters at $z\sim6$ constructed in the same manner as described in Section \ref{sec:mass_model}. 
After the correction of the lensing magnification, 
we obtain an effective survey area of $\sim49$ (2) arcmin$^{2}$ at $L_{\rm [CII]}=1.0\times10^{9}$ (10$^{8}$) $\,{L_{\odot}}$, assuming the line width of FWHM=200 km s$^{-1}$ with the 5$\sigma$ detection limit. 
We then convert the effective survey area to the survey volume, 
based on the frequency setup in the ALCS observations covering the \cii\ line emission at $z=5.974$--6.172 and 6.381--6.602, 
and derive a lower limit of the \cii\ luminosity function at $z=6$. 

In Figure \ref{fig:cii_LF}, we present the number density of \cii\ line emitters at $z=6$, including recent \cii\ line studies at $z>4$ \citep{swinbank2012, matsuda2015, yamaguchi2017, cooke2018, hayatsu2019, yan2020, decarli2020}. 
For comparison, we also present \cii\ luminosity functions from semi-analytical models \citep{popping2016,lagache2018} and from the observed SFR function \citep[SFRF; ][]{smit2016} of optically-selected galaxies. 
For the conversion from SFRF to \cii\ luminosity function, we adopt the \cii--SFR relation of the local star-forming galaxies estimated in \cite{delooze2014}. 
We find that our lower limit estimate is consistent with both the semi-analytical results and SFRF.  
Note that we do not apply any completeness corrections to our lower limit estimate. The incompleteness for strongly lensed sources with large spatial sizes is generally significant due to its low surface brightness \citep[e.g.,][]{bouwens2017, kawamata2018, fujimoto2017}. 
Although the incompleteness largely depends on the assumption of the intrinsic source size, 
this may indicate that the lower limit could be placed still higher and that the faint-end of the \cii\ luminosity function might be close to SFRF. 
Indeed, other constraints from the recent \cii\ line studies are also consistent with SFRF at the bright regime ($L_{\rm [CII]}\gtrsim10^{8.5}\,L_{\odot}$). 
Although the brightest-end ($L_{\rm [CII]}\gtrsim10^{9.2}\,L_{\odot}$ ) of SFRF is smaller than the constraints obtained from the SMG studies \citep{cooke2018}, this is explained by the absence of such dusty-obscured galaxies in the SFRF based on the optically-selected galaxies. 
Therefore, the constraints of the \cii\ luminosity function so far obtained are likely consistent with the prediction from the local SFR--$L_{\rm [CII]}$ relation and the SFRF at $z=6$. 

\subsection{From ISM to Cosmic Scales}
\label{sec:ism-cosmic}
The source plane reconstruction in Section~\ref{sec:morph}  unveils the ISM structure down to a few hundred parsec scales, where we find that \cii\ line is not displaced beyond a $\sim$ 300-pc from the rest-frame UV continuum from the spatially resolved ISM to the whole galaxy. 
In Section~\ref{sec:sfr-cii}, we find that the SFR--$L_{\rm [CII]}$ relations from the spatially resolved ISM to the whole galaxy are consistent with those of local galaxies.  
In Section \ref{sec:cii_LF}, 
we obtain the lower limit at the faintest regime of the $z=6$ \cii\ luminosity function. We find that the prediction from the $z=6$ SFR function and the SFR--$L_{\rm [CII]}$ relation of the local galaxies is consistent with our and previous constraints on the $z=6$ \cii\ luminosity function in the wide $L_{\rm [CII]}$ range. 
Given the unbiased aspect of the ALCS survey and indeed the representative physical properties of \targbg\ among the abundant population of the low-mass regime of $z\sim6$ star-forming galaxies (Section \ref{sec:int_phy}), 
our results may imply that the SFR--$L_{\rm [CII]}$ relation of local star-forming galaxies is universal for a wide range of scales including the spatially resolved ISM, the whole region of the galaxy, and the cosmic scale. 

\section{Summary}
\label{sec:summary}
In this paper, we present the blind detection of a multiply-imaged line emitter behind the massive galaxy cluster RXCJ0600$-$2007 in a cycle-6 ALMA large project of ALMA Lensing Cluster Survey (ALCS). 
The optical--NIR property and our lens model analyses suggest that the emission line is the \cii\ 158 $\mu$m line from a Lyman-break galaxy (LBG) at $z=6.0719\pm0.0004$ behind RXCJ0600-2007. 
We study the relation between the star-formation rate (SFR) and \cii\ line luminosity ($L_{\rm [CII]}$), the morphology, and the kinematics in the spatially resolved interstellar medium (ISM) as well as the whole scale of the LBG, 
and provide a lower limit at the faint-end of the \cii\ luminosity function at $z\sim6$, with help of the gravitational lensing magnification. 
The main findings of this paper are summarized as follows:

\begin{enumerate}
\item We perform  blind line search for the ALCS data cube in RXCJ0600-2007 and identify two bright lines at $\geq8\sigma$ levels at 268.682 $\pm$ 0.011 GHz and 268.744 $\pm$ 0.016 GHz, one of which shows a strongly lensed arc shape. Both lines have optical--NIR counterparts with clear Lyman-break feature at $\sim$9000 ${\rm \AA}$, indicative of the lines corresponding to the \cii\ 158 $\mu$m at $z=6.07$. The optical--NIR spectral energy distribution (SED) analysis shows that probability distributions of their photometric redshifts are in excellent agreement with the \cii\ line redshift, while other possible FIR lines at $z\sim6$ are hard to explain the luminosity ratio between the line and continuum. We thus conclude that these two lines are \cii\ lines. 
\item Our lens models, updated with the latest spectroscopic follow-up results with VLT/MUSE, suggest that these lines arise from a strongly magnified and multiply imaged ($\mu\simeq20-160$)  Lyman-break galaxy (LBG) at $z=6.0719$ with a circularized effective radius of $\sim$0.8 kpc and an intrinsic luminosity in the rest-frame UV $\sim3$ times fainter ($M_{\rm UV}$ = $-19.7^{+0.5}_{-0.4}$) than the characteristic luminosity at this epoch. A sub region of the LBG crosses the caustic line in the source plane and thus stretched into an arc over $\sim6''$ in the image plane, for which the \cii\ line is also significantly detected. Our lens models also predict another two multiple images in this field. We identify the sources at the predicted positions and find that their optical--NIR colors agree with the other multiple images of the LBGs. One of them falls in the ALCS area coverage, where we detect a tentative \cii\ line ($3.0\sigma$) at the same frequency as the other multiple images.  
\item  After the correction of the lensing magnification, the whole of the LBG and its sub region are characterized with $L_{\rm [CII]}$ of  1.1$^{+0.7}_{-0.5}$  $\times10^{8}$ and 0.3$^{+0.1}_{-0.1}$ $\times10^{8}\, L_{\odot}$,  SFR of 5.4$^{+4.5}_{-2.9}$ and 0.8$^{+0.4}_{-0.2}$ $M_{\odot}$ yr$^{-1}$, and stellar mass ($M_{\rm star}$) of 9.6$^{+6.0}_{-4.6}$  $\times10^{8}$ and  2.6$^{+0.2}_{-1.0}$ $\times10^{8}\,M_{\odot}$, respectively. From the whole to sub regions of the LBG, the SFR and $M_{\rm star}$ values falls on the average relation among $z\sim6$ galaxies, indicating that the LBG is an abundant, representative galaxy at this epoch. 
\item The source plane reconstruction resolves the ISM down to $\sim$100--300 pc. The \cii\ line from the whole region of the LBG is co-spatial with the rest-frame UV continuum, while the sub region of the LBG is placed $\sim$1.6 kpc away from the galactic center and bright rest-frame UV clumps. The two-peak morphology observed in the \cii\ line and rest-frame FIR continuum in the arc show a $\sim$1 kpc offset from the luminosity-weighted center of the sub region of the LBG, which likely consists either of a clumpy structure or a smooth disk but stretched into the two-peak morphology due to the perturbation by a foreground galaxy. In these two peaks, the \cii\ line and the rest-frame FIR continuum exhibit the {\it flux ratio anomaly} differently, which suggests that the faint \cii- and FIR-emitting regions are displaced near the caustic.   
\item We find that our results in both whole and sub regions of the LBG fall on the SFR--$L_{\rm [CII]}$ and surface density of SFR ($\Sigma$SFR)--$L_{\rm [CII]}$/SFR relations obtained in local star-forming galaxies. The sub region of the galaxy has a lower $\Sigma$SFR and a higher $L_{\rm [CII]}$/SFR value. This is consistent with the absence of the bright rest-frame UV clumps around the sub region of the LBG that is placed $\sim$1.6 kpc away from the galactic center, where $\Sigma$SFR is expected to be low.   
\item We find that the LBG is classified as a rotation-dominated system based on the full observed velocity gradient and the velocity dispersion of the LBG via the bright \cii\ line emission. The 3D modeling with {\sc $^{\rm 3D}$Barolo} and {\sc galpak3D} provide consistent results for the rotation kinematics that explains the spatial and velocity offsets of the sub region of the LBG. We estimate the dynamical mass of $M_{\rm dyn}=(3\pm1)\times10^{9}\,M_{\odot}$ and obtain the gas fraction of $\sim50$--80\%. 
\item We derive a lower limit on the \cii\ luminosity function at $z=6$. We find that it is consistent with current semi-analytical model predictions. In conjunction with previous ALMA results, we also find that constraints on the \cii\ luminosity function at $z=6$ so far obtained agree with the prediction from the SFR--$L_{\rm [CII]}$ relation of local star-forming galaxies and the SFR function at $z=6$. 
\item With the blind aspect of the ALCS survey and the SFR--$L_{\rm [CII]}$ relations from the sub to whole regions of the LBG, our results may imply that the local SFR--$L_{\rm [CII]}$ relation is universal for a wide range of scales including the spatially resolved ISM, the whole region of the galaxy, and the cosmic scale even up to $z=6$, which we derive in an unbiased manner. 

\end{enumerate}
We thank the anonymous referee for the careful review and
valuable comments that improved the clarity of the paper. 
We thank Justin Spilker and Tanio D\'iaz-Santos for sharing their measurements. We also thank John R. Weaver and Yuchi Harikane for useful comments on the paper and Francesca Rizzo for helpful comments for the kinematic analysis. 
This paper makes use of the ALMA data: ADS/JAO. ALMA \#2018.1.00035.L. 
ALMA is a partnership of the ESO (representing its member states), 
NSF (USA) and NINS (Japan), together with NRC (Canada), MOST and ASIAA (Taiwan), and KASI (Republic of Korea), 
in cooperation with the Republic of Chile. 
The Joint ALMA Observatory is operated by the ESO, AUI/NRAO, and NAOJ. 
This work is based on observations and archival data made with the {\it Spitzer Space Telescope}, which is operated by the Jet Propulsion
Laboratory, California Institute of Technology, under a contract with NASA along with archival data from the NASA/ESA 
{\it Hubble Space Telescope}. This research made also use of the NASA/IPAC Infrared Science Archive (IRSA), 
which is operated by the Jet Propulsion Laboratory, California Institute of Technology, under contract with the National Aeronautics and Space Administration. 
This work was supported in part by World Premier International
Research Center Initiative (WPI Initiative), MEXT, Japan, and JSPS
KAKENHI Grant Number JP18K03693. 
S.F. acknowledges support from the European Research Council (ERC) Consolidator Grant funding scheme (project ConTExt, grant No. 648179) and Independent Research Fund Denmark grant DFF--7014-00017. 
The Cosmic Dawn Center is funded by the Danish National Research Foundation under grant No. 140.
NL acknowledges the Kavli Fundation. 
GBC and KIC acknowledge funding from the European Research Council through the Consolidator Grant ID 681627-BUILDUP.
F.E.B acknowledges supports from ANID grants CATA-Basal AFB-170002, FONDECYT Regular 1190818, and 1200495, and Millennium Science Initiative ICN12\_009. 
IRS acknowledges support from STFC (ST/T000244/1).
KK acknowledges support from the Swedish Research Council and the Knut and Alice Wallenberg Foundation. 

\software{{\sc casa} (v5.4.0; \citealt{mcmullin2007}), {\sc grizli} \citep{brammer2008}, {\sc dendrogram} \citep{goodman2009}, {\sc galfit} \citep{peng2010}, {\sc eazy} \citep{brammer2008}, {\sc scarlet} \citep{melchior2018}, {$^{\rm 3D}${\rm Barolo}}\citep{diteodoro2015}, {\sc galpak3d} \citep{bouche2015}}, {\sc glafic} \citep{oguri2010}, {\sc lenstool} \citep{jullo2007}, {\sc ltm} \citep{zitrin2015}

\clearpage

\appendix

\section{MUSE Spectroscopic Catalog} 
\label{sec:appendix_spec-z}

In Table \ref{tab:muse_list}, 
we summarize the spectroscopic sample from VLT/MUSE (ESO program ID 0100.A-0792, PI: A. Edge) which we use for constraining our lens mass models.

\begin{center}
\begin{longtable}{lcccc}
\caption{MUSE Spectroscopic Catalog}\label{tab:muse_list} \\
\hline
\hline
RELICS ID &  R.A.       &  Dec.     & $z_{\rm spec}$  &  flag              \\ 
          &  deg     &  deg    &                 &                               \\ 
    (1)      &  (2)        &    (3)     &       (4)  &      (5)                   \\ \hline 
467 & 90.0386151 & $-20.1280873$ & 0.0 & 4 \\
474 & 90.0261297 & $-20.1283111$ & 0.4366 & 2 \\
490 & 90.0410860 & $-20.1282533$  & 0.8943 & 3 \\
508 & 90.0352780 & $-20.1293624$  & 0.0 & 4 \\
510 & 90.0366252 & $-20.1292593$ & 0.4230 & 3 \\
514 & 90.0389765 & $-20.1292234$ & 0.4293 & 3 \\
524 & 90.0333696 & $-20.1293114$ & 0.5614 & 3 \\
538 & 90.0258085 & $-20.1298490$ & 1.0270 & 9 \\
539 & 90.0260218 & $-20.1303021$ & 0.4448 & 3 \\
543 & 90.0333594 & $-20.1313998$ & 0.4299 & 3 \\
545 & 90.0346490 & $-20.1312617$ & 0.4316 & 3 \\
547 & 90.0322874 & $-20.1302526$ & 0.4234 & 3 \\
571 & 90.0288919 & $-20.1307349$ & 0.8751 & 3 \\
606 & 90.0317981 & $-20.1320082$ & 0.2662 & 3 \\
620 & 90.0264851 & $-20.1330304$ & 0.3284 & 3 \\
624 & 90.0292841 & $-20.1331843$ & 0.4164 & 3 \\
625 & 90.0286343 & $-20.1325697$ & 0.3449 & 3 \\
626 & 90.0288471 & $-20.1326848$ & 0.3448 & 3 \\
647 & 90.0377894 & $-20.1333237$ & 0.4176 & 3 \\
660 & 90.0400373 & $-20.1375945$ & 0.3843 & 3 \\
662 & 90.0412169 & $-20.1343049$ & 0.384 & 3 \\
684 & 90.0410355 & $-20.1343563$ & 0.3838 & 3 \\
685 & 90.0359084 & $-20.1336945$ & 0.4190 & 2 \\
687 & 90.0328670 & $-20.1339062$ & 0.2298 & 3 \\
697 & 90.0359808 & $-20.1344771$ & 0.4215 & 3 \\
699 & 90.0321831 & $-20.1357916$ & 0.4245 & 2 \\
705 & 90.0332455 & $-20.1367707$ & 0.0 & 4 \\
711 & 90.0349938 & $-20.1360342$ & 0.4332 & 3 \\
724 & 90.0265792 & $-20.1350258$ & 0.7360 & 3 \\
735 & 90.0274334 & $-20.1354704$ & 0.4280 & 3 \\
736 & 90.0257819 & $-20.1357459$ & 0.4294 & 3 \\
737 & 90.0267125 & $-20.1360004$ & 0.4300 & 3 \\
742 & 90.0340260 & $-20.1357916$ & 0.4266 & 3 \\
743 & 90.0376604 & $-20.1357714$ & 0.4369 & 3 \\
754 & 90.0337757 & $-20.1341905$ & 0.4233 & 2 \\
772 & 90.0278737 & $-20.1374801$ & 0.4307 & 3 \\
779 & 90.0326653 & $-20.1350609$ & 0.4305 & 3 \\
786 & 90.0429917 & $-20.1367709$ & 5.4589 & 9 \\
791 & 90.0281193 & $-20.1381123$ & 0.5089 & 3 \\
792 & 90.0335009 & $-20.1349955$ & 0.4304 & 3 \\
801 & 90.0394961 & $-20.1371561$ & 4.5043 & 3 \\
802 & 90.0348279 & $-20.1356744$ & 0.4319 & 3 \\
806 & 90.0364247 & $-20.1377419$ & 0.4295 & 3 \\
814 & 90.0414618 & $-20.1358813$ & 0.0 & 4 \\
823 & 90.0317133 & $-20.1382886$ & 0.4276 & 3 \\
836 & 90.0264868 & $-20.1381907$ & 0.0 & 4 \\
860 & 90.0365781 & $-20.1393201$ & 0.4177 & 3 \\
862 & 90.0279194 & $-20.1393703$ & 0.4321 & 3 \\
863 & 90.0278743 & $-20.1397608$ & 0.4296 & 3 \\
870 & 90.0430009 & $-20.1395918$ & 0.4392 & 3 \\
871 & 90.0425305 & $-20.1399135$ & 0.3825 & 3 \\
886 & 90.0419909 & $-20.1399491$ & 0.4462 & 2 \\
887 & 90.0340961 & $-20.1398346$ & 0.4315 & 3 \\
899 & 90.0295615 & $-20.1407949$ & 0.4317 & 3 \\
900 & 90.0296018 & $-20.1404579$ & 0.4323 & 3 \\
938 & 90.0430509 & $-20.1413250$ & 0.4197 & 3 \\
941 & 90.0343581 & $-20.1391201$ & 0.4195 & 3 \\
956 & 90.0355945 & $-20.1419747$ & 0.4240 & 2 \\
957 & 90.0387747 & $-20.1421929$ & 0.4314 & 3 \\
962 & 90.0306151 & $-20.1429810$ & 0.0866 & 3 \\
963 & 90.0303906 & $-20.1430867$ & 0.0866 & 3 \\
964 & 90.0304180 & $-20.1428506$ & 0.0866 & 3 \\
965 & 90.0304371 & $-20.1427616$ & 0.0866 & 3 \\
966 & 90.0299297 & $-20.1421244$ & 0.0866 & 3 \\
967 & 90.0300655 & $-20.1424500$ & 0.0866 & 3 \\
973 & 90.0349638 & $-20.1425005$ & 0.4255 & 3 \\
974 & 90.0350788 & $-20.1429515$ & 0.4313 & 3 \\
990 & 90.0258089 & $-20.1429104$ & 0.5491 & 3 \\
1000 & 90.0296362 & $-20.1430670$ & 0.4305 & 3 \\
1003 & 90.0384290 & $-20.1430473$ & 0.5479 & 9 \\
1024 & 90.0337998 & $-20.1436577$ & 2.7722 & 3 \\
1025 & 90.0340880 & $-20.1437083$ & 2.7723 & 3 \\
1027 & 90.0351249 & $-20.1438250$ & 2.7723 & 9 \\
1029 & 90.0356487 & $-20.1438539$ & 2.7725 & 3 \\
900001 & 90.0424970 & $-20.1364998$ & 3.5238 & 3 \\
900003 & 90.0283847 & $-20.1376962$ & 5.4067 & 9 \\
\hline
\end{longtable}
\end{center}
{\footnotesize
(1) ID from the RELICS public catalogue of hlsp\_relics\_hst\_wfc3ir\_rxc0600-20\_multi\_v1\_cat.txt\footnote{https://relics.stsci.edu/}.  
IDs starting with 900 are MUSE detections with no counterpart in the mentioned catalogue.
(2) Observed right ascension in degrees. 
(3) Observed declination in degrees. 
(4) MUSE spectroscopic redshift. 
(5) Redshift quality flag. 2: likely, 3: secure measurement, 9: single line measurement, and 4: field stars. 
}

\section{Two-peak Morphology in $z$6.1/6.2} 
\label{sec:appendix_sim}

To check the possibility that the two-peak morphology of the \cii\ line in $z$6.1/6.2 is caused by the noise fluctuation boosted by the underlying diffuse emission \citep[e.g.,][]{hodge2016}, 
we perform a mock observation with the CASA task {\sc simobserve} towards $z$6.1/6.2 in the same manner as \cite{fujimoto2020a}. 
Here we assume the single elliptical Gaussian for the \cii\ line surface brightness distribution of $z$6.1/6.2 based on the {\sc imfit} results in the $uv$-tapered map (Section \ref{sec:line_prop}).  
We then obtain the visibility data set through {\sc simobserve} and produce the natural-weighted velocity-integrated map of the \cii\ line. 
We repeat the mock observation to producing the map 1,000 times. 
Given that the spatial offset of $\sim2\farcs0$ and the significance levels of 8.2$\sigma$ and 5.4$\sigma$ between the two peaks in $z$6.1/6.2, we then search multiple positive peaks that are located with spatial offsets of less than $2\farcs5$ and detected at $\geq5.4\sigma$ levels, utilizing SExtractor version 2.5.0 \citep{bertin1996}. 
We identify 7 out of 1,000 maps have the multiple peaks that meet the above criteria. 
These results indicate that the two-peak morphology of the \cii\ line in $z$6.1/6.2 might be caused by the noise fluctuation with a probability of $\sim$ 0.7\%. 
Note that we find that all multiple peaks identified in the 7 maps show their flux ratios almost identical, which is different from the two peaks observed in $z$6.1/6.2 (ratio $\sim$ 8:5). 
 This indicates that the close separation as well as the flux ratio of the two peaks observed in $z$6.1/6.2 is hardly explained by the noise fluctuation.   
In fact, we identify only 1 out of 1,000 maps that has a flux ratio of multiple peaks similar to the two peaks in $z$6.1/6.2, but with the spatial offset of $4\farcs7$. 
Therefore, we conclude that the possibility of the noise fluctuation is negligible in the two-peak \cii\ line morphology of $z$6.1/6.2. 

\section{Optical--NIR Photometry} 
\label{sec:appendix_photo}

We adopt separate strategies for extracting robust photometry for the four lensed images as described below to account for the crowded cluster field and varying degrees of extended source morphology.  
In general, we model the full IRAC mosaics using a strategy similar to that of \cite{merlin2015},  where we use image thumbnails of each source 
and neighbors taken from the high-resolution {\it HST}/WFC3 F160W image and knowledge of the WFC3 and IRAC point spread functions (PSFs) to model the low-resolution IRAC image.  

\subsection{Images $z$6.3 and $z$6.5}

The sources of interest in these images are relatively bright and fairly well separated from their nearest bright (projected) neighbors (Figure \ref{fig:fig1}).  We measure aperture flux densities in each of the {\it HST} filters using fixed $D=0\farcs7$ apertures centered on the source of interest to define the colors.  To determine the overall flux normalization, we model the source morphology of the lensed image and nearby neighbors using the non-parametric morphological fitting code {\sc Scarlet} \citep{melchior2018}. 
All of the WFC3/IR images (and their PSFs) are used to constrain the {\sc Scarlet} morphological model.  We scale all of the {\it HST} aperture measurements $F_{ap, \mathrm{i}}$ by the aperture correction $F_{S, \mathrm{F160W}} / F_{ap, \mathrm{F160W}}$, where $F_{S, \mathrm{F160W}}$ is the integral of the {\sc Scarlet} model evaluated in the F160W filter and $F_{S, \mathrm{F160W}}$ is the aperture measurement in that filter.  The photometric uncertainties are measured in the same apertures on the inverse variance image in each filter. 
For the IRAC flux densities of these images, we subtract all modeled sources other than the source of interest and perform aperture photometry on this cleaned image using $D=3\farcs0$ apertures, which we correct to the same ``total'' scale as for {\it HST} using aperture corrections of 1.6 and 1.7 for channels 1 and 2, respectively, that were derived from a separate bright, isolated source in the field.

\subsection{Extended arc image $z$6.1/6.2}
\label{sec:appendix_phot_z6.1}

This image is a highly elongated arc extending over $\approx$6\,arcsec coincident with a foreground compact source in the center (Figure \ref{fig:fig1}).  Here, we model both overlapping sources in the F160W image as parametric Sersic profiles using the {\sc galfit} software \citep{peng2010}.  For the photometry of the lensed arc and foreground image in the WFC3/IR filters, we fit for the relative normalizations of the two Sersic components convolved with the appropriate PSFs.  For IRAC, we convolve the model Sersic profiles with the IRAC PSF and fit for the normalization of the source of interest and all neighboring sources in the least-squares optimization.  As for HST, the normalization of the scaled morphological components is adopted as the photometric measurement without additional aperture corrections.  For the optical images where the arc is not readily visible, we measure an aperture flux density and its associated uncertainty within a large rectangle aperture approximately $1\farcs2\times3\farcs0$.

Note that the de-blended color of the foreground object is similar to the color of cluster members, and the best-fit SED shows the photometric redshift at $0.57^{+0.14}_{-0.17}$ which is close to the cluster redshift at $z=0.43$ (see also Laporte et al. submitted). 
Although this suggests the foreground to be one of the cluster members, we do not include it in our fiducial mass model due to potential systematics in the de-blending process. The detail contribution of the foreground object to morphology and magnification factors of $z$6.1/6.2 (Section \ref{sec:morph}) must be investigated after we obtain the spectroscopic redshift of the foreground object.

\subsection{Faint image $z$6.4}
\label{sec:appendix_phot_z6.4}

The final faint image of $z$6.4 is close to the cluster core and the BCG.  Although it is not deblended as a separate source in our original photometric catalog (and associated IRAC model), a source is readily apparent in the F160W image (Figure \ref{fig:fig1}).  
We estimate photometry of this image by placing fixed $D=0\farcs7$ and $D=3\farcs0$ apertures centered on the F160W position in the {\it HST} and IRAC filter mosaics, respectively, and scale these measurements by aperture corrections derived for point sources.

\section{\cii\ Spectra of $\lowercase{z}6.1$ and $\lowercase{z}6.2$} 
\label{sec:appendix_z6.1-6.2}

\begin{figure}
\begin{center}
\includegraphics[trim=0cm 0cm 0cm 0cm, clip, angle=0,width=0.5\textwidth]{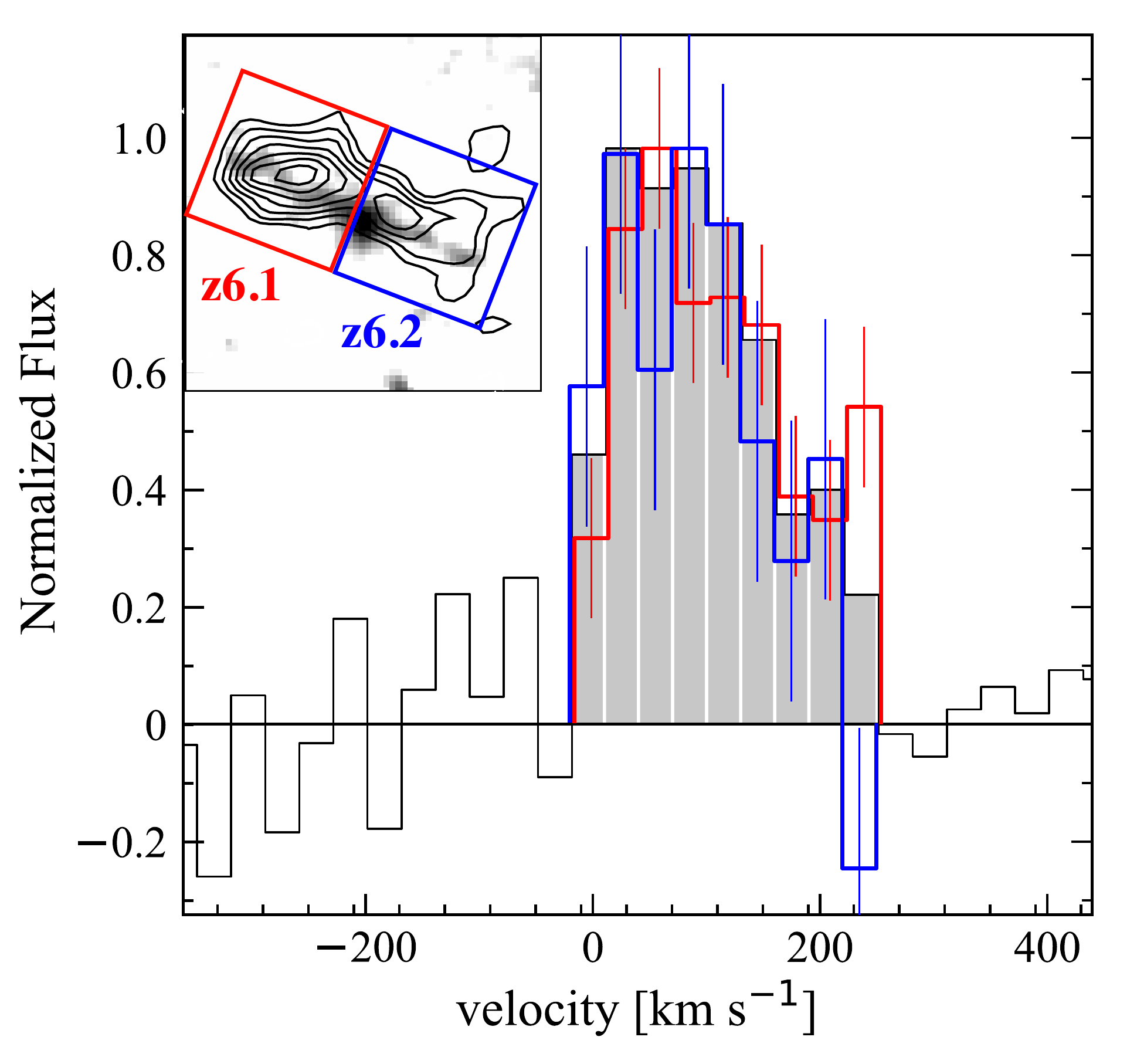}
\end{center}
\vspace{-0.2cm}
 \caption{
Zoom-in \cii\ line spectra of $z6.1$ and $z$6.2 as a function of velocity with respect to the frequency peak of $z$6.3. 
The inset panel shows the same image cutout as the middle panel of Figure \ref{fig:fig1} for $z6.1$/6.2. 
The red and blue squares denote apertures that are used to produce the \cii\ spectra for $z6.1$ and $z6.2$, respectively. 
The black line and the grey shade indicate the integrated \cii\ line spectrum of $z6.1$/6.2 
and the \cii-detected channels, respectively. 
The red and blue lines present the \cii\ spectra for $z6.1$ and $z6.2$, respectively, that are normalized to the peak of the integrated one. 
\label{fig:z6.1-6.2}}
\end{figure}

As a sanity check of our interpretation that one of the \cii\ line emitters consists of a pair of multiple images of $z6.1$ and $z6.2$, 
we compare \cii\ spectra between $z6.1$ and $z6.2$. 
In Figure \ref{fig:z6.1-6.2}, we show the \cii\ spectra of $z6.1$ (red line) and $z6.2$ (blue line) that are normalized to the peak of $z$6.1/6.2 (black line).
We find that $z6.1$ and $z6.2$ have \cii\ line profiles consistent with each other within the errors, 
which agrees with our interpretation of $z6.1$ and $z6.2$ being a pair of multiple images.

\section{\cii\ rotation modeling} 
\label{sec:appendix_kin}

In Section \ref{sec:kin}, we find that $z6.3$ is likely a rotation-dominated system. In Table \ref{tab:3dmodel}, we summarize the 3D modeling results for the \cii\ line around $z6.3$ with $^{\rm 3D}${\sc barolo} \citep{diteodoro2015} and {\sc galpak3d} \citep{bouche2015}. 
For $^{\rm 3D}${\sc barolo}, because the ALMA beam has a half-width-at-half-maximum (HWHM) of  $\sim 0\farcs45$ at the \cii\ line frequency along the orientation of the velocity gradient, we adopt three ($\sim$ 1.4/0.45) annuli with the width of $0.''45$ for the tilted ring fitting algorithm. We use the THRESHOLD mask with the 2$\sigma$ limit for the data cube, and the spatial center, systemic velocity, rotation velocity ($v_{\rm rot}$), velocity dispersion ($\sigma_{\rm vel}$), position angle (PA), and inclination (incl.) are used as free parameters in the fitting. The errors are estimated based on the minimization algorithm in a Monte Carlo approach. 
For {\sc galpak3d}, we adopt the exponential-disk for the flux profile, the Gaussian for the thickness profile, and the $\tanh$ formalization of $V_{\rm max}\times \tanh(r/r_{\rm V})$ for the rotation curve, where $V_{\rm max}$ and $r_{\rm V}$ are the maximum velocity and the turnover radius, respectively. 
In place of the mask, we use a cutout data cube by $3\farcs8\times3\farcs8$ and [$-120$: +120] km~s$^{-1}$ for the fitting.  We set the maximum iteration number of 20,000.  
The spatial center, systemic velocity, flux, $r_{\rm e}$, PA, incl., $r_{\rm V}$, $V_{\rm max}$, and $\sigma_{\rm vel}$ are used as free parameters.  The errors are evaluated based on a Markov chain Monte Carlo approach.

\tcb{
\begin{table}[h]
\begin{center}
\caption{3D Modeling Results}
\label{tab:3dmodel}
\begin{tabular}{lcccc}
\hline 
\hline
& \multicolumn{3}{c}{$^{\rm 3D}${\sc barolo}}& {\sc galpak3d} \\ \hline
     &  Ring1        &     Ring2      &    Ring3    &  Exponential disk  \\  \hline
radius$^{a}$ [$''$] & 0--0.45 & 0.45--0.9 & 0.9--1.35 &   0.55 $\pm$ 0.02   \\
$v_{\rm LOS}$$^{b}$ [km s$^{-1}$]& 22$^{+5}_{-4}$ & 76$^{+22}_{-22}$ & 62$^{+4}_{-3}$ & 62 $\pm$ 4 \\
$\sigma_{\rm vel}$ [km s$^{-1}$]  & 46$^{+9}_{-8}$ & 54$^{+11}_{-11}$ & 59$^{+7}_{-7}$ & 55 $\pm$ 3\\
incl. [deg]                         & 56 & 57 & 59   & 57 $\pm$ 3 \\
PA$^{c}$  [$^{\circ}$]               & 36 & 32 & 27   & 21 $\pm$ 3 
\\
\hline
\end{tabular}
\end{center}
$a$ Inner and outer radii of the tilted rings for $^{\rm 3D}${\sc barolo} and effective radius for {\sc galpak3d}. The turnover radius is estimated to be $0\farcs62 \pm 0\farcs11$ in the $\tanh$ formalization for the rotation curve in {\sc galpak3d}. \\
$b$ Line-of-sight projected rotation velocity. We present $V_{\rm max}$ for {\sc galpak3d}. \\
$c$ We follow the definition of PA as the orientation from the y-axis in anti-clockwise, which is different from that of the original output of $^{\rm 3D}${\sc barolo}. 
\end{table}
}

\bibliographystyle{apj}
\bibliography{apj-jour,reference}

\clearpage

\end{document}